\documentclass[11pt,preprint]{aastex}
\usepackage{graphics,color}

\shorttitle{Intrinsic Shape of 2MASS Spirals}
\shortauthors{Ryden}

\begin{document}

\title{The Intrinsic Shape of Spiral Galaxies\\
in the 2MASS Large Galaxy Atlas}
\author{Barbara S. Ryden}
\affil{Department of Astronomy, The Ohio State University}
\affil{140 W. 18th Avenue, Columbus, OH 43210\\
     ryden@astronomy.ohio-state.edu
}

\begin{abstract}

The apparent shapes of spiral galaxies in the 2-Micron
All Sky Survey Large Galaxy Atlas are used to constrain
the intrinsic shapes of their disks. When the distribution
of apparent axis ratios is estimated using a nonparametric
kernel method, the shape distribution is inconsistent
with axisymmetry at the 90\% confidence level in the $B$
band and at the 99\% confidence level in the $K_s$ band.
If spirals are subdivided by Hubble type, the late-type
spirals (Sc and later) are consistent with axisymmetry,
while the earlier spirals are strongly inconsistent with
axisymmetry. The distribution of disk ellipticity can be
fitted adequately with either a Gaussian or a lognormal
distribution. The best fits for the late spirals imply
a median ellipticity of $\epsilon \approx 0.07$ in the
$B$ band and $\epsilon \approx 0.02$ in the $K_s$ band.
For the earlier spirals, the best fits imply a median
ellipticity of $\epsilon \approx 0.18$ in the $B$ band
and $\epsilon \approx 0.30$ in the $K_s$ band. The
observed scatter in the Tully-Fisher relation, for both
late and early spirals, is consistent with the disk
ellipticity measured in the $B$ band. This indicates that
excluding spirals of Hubble type earlier than Sc will minimize
the intrinsic scatter in the Tully-Fisher relation used as
a distance indicator.

\end{abstract}


\section{Introduction}
\label{intro}

The intrinsic shapes of disks in spiral galaxies, although
difficult to determine, are useful to know. In particular,
the ellipticity of a disk in its outer region reflects the
nonaxisymmetry of the potential in which it exists, and thus
provides information about the shape of dark matter halos.
Non-circular disks produce scatter in the Tully-Fisher relation
between rotation velocity and absolute magnitude \citep{tu77,fr92};
therefore, knowledge of the disk ellipticity helps in understanding
the origin of scatter in the Tully-Fisher relation. In particular,
the usefulness of the Tully-Fisher relation as a distance indicator
would be enhanced for a population of disk galaxies with small
intrinsic ellipticity. If the shape of a disk
is approximated as an ellipsoid with principal axes
of length $a \geq b \geq c$, the disk shape can be described
by two parameters: the dimensionless disk thickness, $\gamma
\equiv c/a$, and the disk ellipticity, $\epsilon \equiv
1 - b/a$. The simplest approximation, that disks are
infinitesimally thin and perfectly circular ($\gamma = 0$,
$\epsilon = 0$), is occasionally useful; for many purposes,
however, it is useful to have a more accurate estimate of the
shapes of disks.

The thickness of disks can be determined from catalogs of
edge-on spiral galaxies, such as the Flat Galaxy Catalogue
\citep{ka93} and the Revised Flat Galaxy Catalogue \citep{ka99}.
Some superthin galaxies have axis
ratios as small as $\gamma \approx 0.05$ \citep{go81}.
The study of \citet{gu92}, using images of edge-on spirals
from blue Palomar Sky Survey plates, found that $\log q$
had a mean value of $-0.95$, where $q$ is the apparent
axis ratio. The mean disk
thickness depends on the spirals' morphological type
\citep{bo83}; \citet{gu92} found that the mean value of $\log q$ ranges
from $-0.7$ for edge-on Sa galaxies to $-1.1$ for edge-on Sd galaxies.
The disk thickness of an individual spiral galaxy also depends
on the wavelength at which it is observed. Edge-on spirals have
strong vertical color gradients, becoming much redder with
increasing height above the midplane \citep{da02}. This
gradient reflects the transition from the thin disk, with
its young blue stellar population, to the thick disk, with
its older redder population. As a consequence, $q$ of an
edge-on galaxy becomes larger at longer wavelengths. For
example, the galaxies in the \citet{da00} sample were chosen
to have $q \leq 0.125$ at blue wavelengths; in deep $R$-band
images, the typical axis ratio increases to $q \approx 0.23$
\citep{da02}. Similarly, \citet{mi03} found that edge-on spirals
that have $q \approx 0.17$ in blue images have $q \approx 0.35$
in two-micron images.

Estimates of the distribution of disk ellipticity $\epsilon$ can be
made from the distribution of apparent axis ratios $q$ for a large
population of randomly oriented spiral galaxies. The signature of
non-zero ellipticity, in this case, is a scarcity of nearly circular
($q \approx 1$) spiral galaxies. \citet{bi81}, using spiral galaxies
from the Second Reference Catalogue of Bright Galaxies \citep{de76},
concluded that late-type spirals were better fitted by slightly
elliptical disks ($\epsilon \approx 0.1$) than by perfectly circular
disks. More recent studies \citep{gr85, la92, fa93, al02, ry04}
have confirmed the ellipticity of disks, and have permitted estimates
of the distribution function for intrinsic ellipticity. The distribution of
$\epsilon$ is commonly fitted with a half-Gaussian peaking at $\epsilon = 0$.
Values of $\sigma_\epsilon$, in this case, range from $\sigma_\epsilon = 0.12$
\citep{la92}, through $\sigma_\epsilon = 0.13$ \citep{fa93}, to
$\sigma_\epsilon = 0.21$ \citep{ry04}.

Kinematic studies can also provide an estimate of the ellipticity
of the gravitational potential in the disk plane. In an elliptical
potential with small ellipticity $\epsilon_\Phi$, the closed stellar
orbits will themselves be ellipses \citep{bi78}. When the
resulting elliptical stellar disk is seen in projection, the
isophotal principal axes will be misaligned with the kinematic
principal axes. Because of the misalignment, there will be a
velocity gradient along the isophotal minor axis proportional
to $\epsilon_\Phi \sin 2 \phi$, where $\phi$ is the azimuthal
viewing angle relative to the long axis of the potential
\citep{fr92}. Measurement of the kinematic misalignment for
a sample of 9 disk galaxies yielded an average ellipticity of
$\epsilon_\Phi \approx 0.044$ \citep{sc97,sc98}. A similar
analysis of gas kinematics in 7 disk galaxies by \citet{be99}
found an average ellipticity of $\epsilon \approx 0.059$
for the gas orbits. \citet{an02}, using the
method of \citet{an01}, combined two-dimensional
kinematic and photometric data for a sample of 28 nearly
face-on disk galaxies; they found an average halo ellipticity
of $\epsilon_\Phi \approx 0.054$ and an average disk ellipticity
of $\epsilon \approx 0.076$ for their sample. Taking into account
the selection criteria of \citet{an02}, the implied median
disk ellipticity is $\epsilon \approx 0.10$ \citep{ry04}.

Much of what we know about the ellipticity of disks thus comes
from looking at galaxies with low inclination; much of
what we know about disk thickness comes from looking at
galaxies with high inclination. In principle, both ellipticity
and thickness can be determined by looking at a sample of
disk galaxies with random inclinations \citep{fa93}. However,
at visible wavelengths, thanks to the effects of dust,
it's difficult to create a sample in which the
disk inclinations are completely random.
Because disk galaxies are not fully transparent,
a flux-limited sample will be deficient in nearly
edge-on galaxies \citep{hu92}.
Because disk galaxies are not fully opaque, an angular
diameter-limited sample will have an excess of nearly
edge-on galaxies \citep{hu92,ma03}.

To minimize the effects of dust, I will be examining
the apparent shapes of disk galaxies in the Two Micron
All-Sky Survey (2MASS). In \S\ref{sec-sample}, I extract
a randomly inclined sample of spiral galaxies from the
2MASS Large Galaxy Atlas. In \S\ref{sec-axisym}, I make
both nonparametric and parametric estimates of $f(\gamma)$,
the distribution of disk thickness, assuming the disks
are axisymmetric. Separate estimates are made for
early-type spirals and for late-type spirals. In
\S\ref{sec-nonaxi}, I make nonparametric estimates of
$f(\epsilon)$, the distribution of disk ellipticity
for nonaxisymmetric disks. Both Gaussian and lognormal
functions are found to give satisfactory fits to $f(\epsilon)$.
In \S\ref{sec-disc}, I examine the influence of the deduced
disk ellipticity on the scatter of the Tully-Fisher relation,
and consider the possible origins of disk ellipticity for
early-type and late-type spirals.

\section{The Galaxy Sample}
\label{sec-sample}

The 2MASS Large Galaxy Atlas (LGA) is a catalog of galaxies with
large angular size as seen in the near-infrared \citep{ja03}.
The LGA provides an essentially complete sample of galaxies with
$R_{20} \geq 60 \arcsec$, where $R_{20}$ is the semimajor
axis of the $K_s$ band $20 {\rm\,mag} {\rm\,arcsec}^{-2}$ isophote.
Observations in the $K_s$ band, with an effective wavelength
of $2.2 \mu{\rm m}$, have two advantages for the study of galaxy
morphology. First, the effects of dust are minimized. Second,
the light in the $K_s$ band emphasizes the smoother distribution
of population II stars rather than the patchier distribution
of population I stars, which tend to lie along spiral arms.
The LGA galaxy images consist of individual 2MASS co-add images
mosaicked together. The typical $1 \sigma$ background noise in
a single $K_s$ co-add image is $\sim 20 {\rm\,mag}{\rm\,arcsec}^{-2}$;
the point spread function FWHM varies from 2 to 3 arcseconds,
depending on seeing \citep{ja03}. The on-line Large Galaxy Atlas\footnote{
See http://irsa.ipac.caltech.edu/applications/2MASS/LGA/}, version 2.0,
contains 620 galaxies. To create a sample of spiral galaxies, I
selected those galaxies with morphological classification SA,
SAB, SB, or S. Galaxies labeled as `pec' were excluded, as were
S0 galaxies. The resulting sample contains 383 spiral
galaxies.

The apparent shape of each galaxy in the $K_s$ band is estimated
by \citet{ja03} from the axis ratio of the $3 \sigma$ intensity
isophote, corresponding to a surface brightness of $\mu_K \approx
18.8 {\rm\,mag}{\rm\,arcsec}^{-2}$; I call this axis ratio $q_K$.
For purposes of comparison, the apparent axis ratio $q_B$ in the $B$
band was estimated by taking the axis ratio of the $\mu_B =
25 {\rm\,mag}{\rm\,arcsec}^{-2}$ isophote from the LEDA
database. Because the 2MASS Large Galaxy Atlas is an angular
diameter-limited sample and spiral galaxies are largely
transparent in the $K_s$ band, the Large Galaxy Atlas contains
an excess of high-inclination disks with small $q_K$. For instance, of
the 30 LGA galaxies with the largest value of $R_{20}$, 10 are spiral
galaxies with $q_K < 0.33$.

To account for inclination effects, we must estimate what
the isophotal radius of each galaxy would be if we saw
it face-on. Suppose, to begin, that the isoluminosity surfaces
of the galaxy are adequately approximated as similar, concentric,
coaxial ellipsoids, given by the formula
\begin{equation}
m^2 = x^2 + y^2 / \beta^2 + z^2 / \gamma^2 \ ,
\end{equation}
where $1 \geq \beta \geq \gamma$. The galaxy is viewed from
an angle $(\theta,\phi)$, where $\theta$ is the polar angle
measured from the positive $z$ axis and $\phi$ is the azimuthal
angle measured from the positive $x$ axis. Assuming perfect
transparency, the isophotes of the galaxy will be ellipses,
given by the formula
\begin{equation}
M^2 = D X^2 + E Y^2 \ .
\end{equation}
If I define, following \citet{bi78},
\begin{equation}
A(\theta,\phi) = \cos^2 \theta (\beta^2 \sin^2\phi + \cos^2 \phi ) +
\gamma^2 \sin^2 \theta \ ,
\end{equation}
\begin{equation}
B(\theta,\phi) = (1-\beta^2) \cos \theta \sin 2 \phi \ ,
\end{equation}
\begin{equation}
C(\phi) = \sin^2 \phi + \beta^2 \cos^2 \phi \ ,
\end{equation}
and
\begin{equation}
f(\theta,\phi) = \gamma^2 \sin^2 \theta (\beta^2 \cos^2 \phi +
\sin^2 \phi ) + \beta^2 \cos^2 \theta \ ,
\end{equation}
then
\begin{equation}
D (\theta,\phi) = {1 \over 2 f} [ A + C - \sqrt{ (A-C)^2 + B^2 } ]
\label{eq:d}
\end{equation}
and
\begin{equation}
E (\theta,\phi) = {1 \over 2 f} [ A + C + \sqrt{ (A-C)^2 + B^2 } ]
\label{eq:e}
\end{equation}
\citet{ja03} find that LGA galaxies are well fitted, outside their
central nucleus or core, by a S\'ersic profile,
\begin{equation}
\mu (X) = \mu_d \exp \left[ - ( X / \alpha )^{1/n} \right] \ . 
\label{eq:ser}
\end{equation}
The functional form adopted by \citet{ja03}, in which $\alpha$
is a scale radius, is readily converted to the more usual form,
\begin{equation}
\mu (X) = \mu (R_e) \exp \left( - b_n [ ( X / R_e )^{1/n} - 1 ] \right) \ ,
\label{eq:ser2}
\end{equation}
in which $R_e$ is the effective radius, or half-light radius. The
conversion from scale length $\alpha$ to effective radius $R_e$ is
$R_e = (b_n)^n \alpha$, where an adequate approximation to $b_n$ is
\citep{ci99}
\begin{equation}
b_n \approx 2 n - {1 \over 3} + {4 \over 405 n} + {46 \over 25515 n^2} \ .
\end{equation}
For the spiral galaxies in the LGA, the best fitting profiles are
usually close to exponential; 90\% of the spiral galaxies have
$0.7 \leq n \leq 1.4$ in the $K_s$ band. In this range of $n$, the
effective radius ranges from $R_e = 1.06 \alpha$ (when $n = 0.7$) to
$R_e = 3.55 \alpha$ (when $n = 1.4$).

If a galaxy has a surface brightness profile given by equation~(\ref{eq:ser}),
then if it were viewed face-on (that is, from along
the $\theta = 0$ axis) its profile would be
\begin{equation}
\mu (X) = \mu_d^o \exp \left[ - ( X / \alpha^o )^{1/n} \right]  \ .
\label{eq:ser0}
\end{equation}
Here, $\alpha^o$ is the scale length along the apparent
long axis and $\mu_d^o$ is the central surface brightness,
as seen from the angle $\theta = 0$.
Let $R_{20}^o$ be the value of $X$ at which the surface
brightness falls to the specified isophotal brightness
$\mu_{20} = 20 {\rm\,mag}{\rm\,arcsec}^{-2}$. For the profile
of equation~(\ref{eq:ser0}),
\begin{equation}
R_{20}^o = \alpha^o \left[ \ln ( \mu_d^o / \mu_{20} ) \right]^n \ .
\end{equation}
If we actually view the disk galaxy from an angle
$(\theta,\phi)$, the central surface brightness of the disk is
\begin{equation}
\mu_d (\theta,\phi) = \mu_d^o \beta \sqrt{ D(\theta,\phi)
E (\theta,\phi) } 
\end{equation}
and the scale length along the apparent long axis is
\begin{equation}
\alpha (\theta,\phi) = \alpha^o \big/ \sqrt{D(\theta,\phi)} \ .
\end{equation}
The isophotal radius $R_{20}$, at which the surface brightness
on the long axis falls to $\mu_{20}$, is
\begin{equation}
R_{20} (\theta,\phi) = \alpha  \left[ \ln ( \mu_d / \mu_{20} ) \right]^n \ .
\end{equation}
The apparent axis ratio of the disk, seen from the angle
$(\theta, \phi)$, is $q = \sqrt{D/E}$. Thus, the
face-on isophotal radius, $R_{20}^o$, is related
to the observed values of $R_{20} (\theta,\phi)$ and
$\alpha (\theta,\phi)$ by the relation
\begin{equation}
R_{20}^o = \sqrt{D} R_{20} \left[ 1 + ( \alpha / R_{20} )^{1/n}
\ln \left( {q \over \beta D} \right) \right]^n \ .
\label{eq:incor}
\end{equation}
Although $R_{20}$, $\alpha$, and $q$ can be measured for
any particular galaxy seen in projection, $\beta$ and
$D(\theta,\phi)$ cannot. However, if the disk is axisymmetric,
then $\beta = D = 1$, the scale length $\alpha$ is independent
of viewing angle, and the face-on isophotal radius can
be deduced from the observed properties:
\begin{equation}
R_{20}^o = R_{20} \left[ 1 + ( \alpha / R_{20} )^{1/n} \ln q \right]^n \ .
\label{eq:axis}
\end{equation}
For all $n > 0$, an axisymmetric disk appears smallest when
seen face-on ($R_{20}^o \leq R_{20}$).

\section{Shape of Axisymmetric Disks}
\label{sec-axisym}

If we assume that spiral galaxies are axisymmetric, then the
analysis of the preceding section suggests a way of assembling
an inclination-independent subsample of disk galaxies from the
2MASS Large Galaxy Atlas. If the LGA contains all galaxies
with $R_{20} > 60 \arcsec$, then, assuming axisymmetry,
it will contain all galaxies with $R_{20}^o > 60 \arcsec$.
Thus, from the tabulated values for
$R_{20}$, $\alpha$, $n$, and $q$ \citep{ja03}, I compute
$R_{20}^o$ for each spiral galaxy, using
equation~(\ref{eq:axis}). If $R_{20}^o > 60 \arcsec$, the
spiral is included in the inclination-independent subsample.

In this way, I find that 193 spiral galaxies in
the 2MASS Large Galaxy Atlas have $R_{20}^o > 60 \arcsec$.
Of these galaxies, 130 have a Hubble type earlier
than Sc (14 S0/a, 6 Sa, 28 Sab, 40 Sb, and 42 Sbc); I'll
call these galaxies the ``early spirals''. The remaining
63 spirals have a Hubble type of Sc or later (36 Sc, 18 Scd,
8 Sd, and 1 Sm); I'll call these the ``late spirals''.
If I had not done the inclination correction to find $R_{20}^o$,
I would have found 161 early spirals and 117 late spirals
with $R_{20} > 60 \arcsec$.

\begin{figure}
\plotone{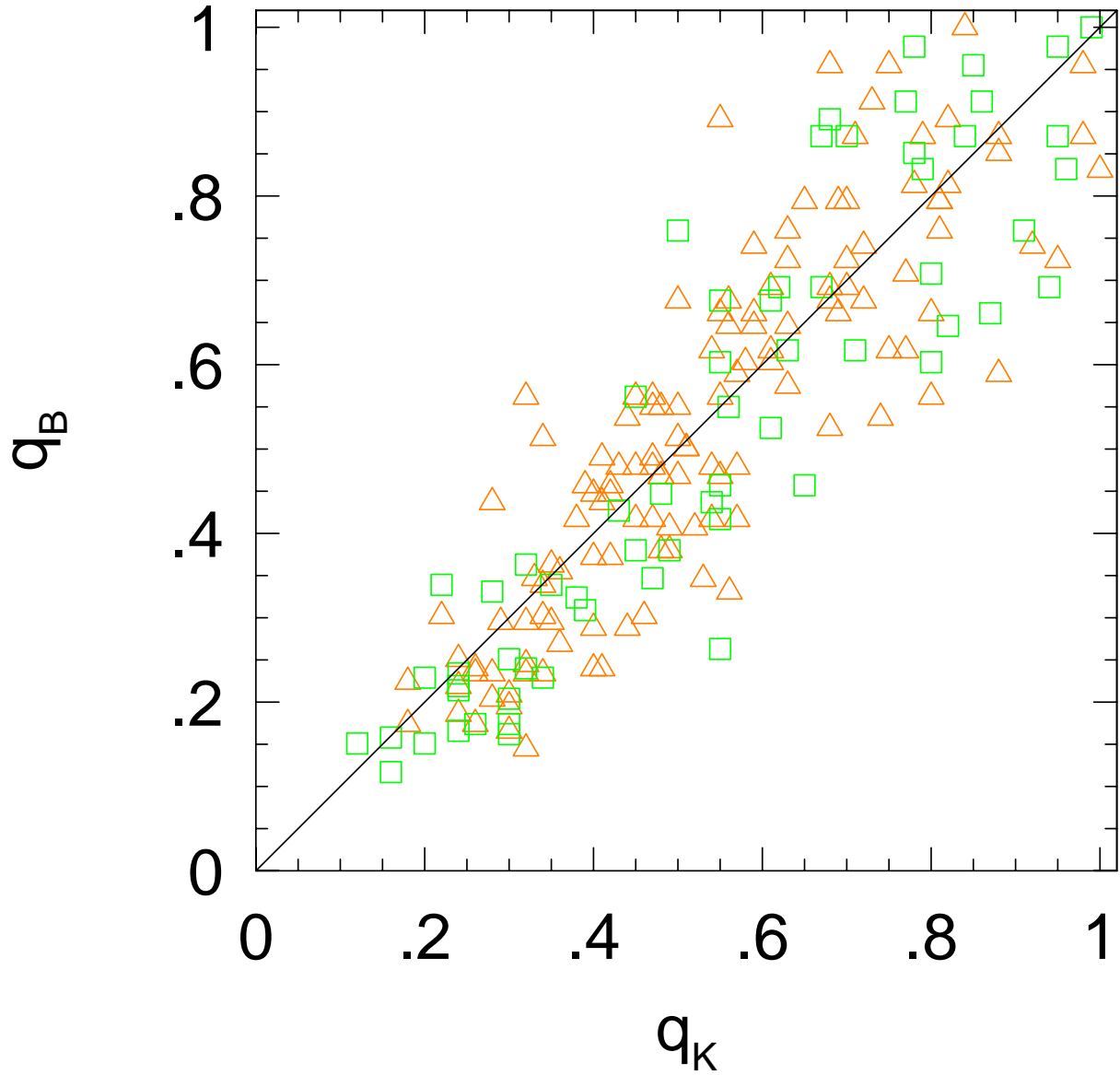}
\caption{Axis ratio of the 25 mag/arcsec$^2$ isophote in
the $B$ band versus the axis ratio of the $3\sigma$ isophote
in the $K_s$ band. Squares indicate late spirals (Sc or later);
triangles indicate early spirals.
}
\label{fig:qkqb}
\end{figure}

The apparent axis ratios $q_K$ and $q_B$ for the 193
LGA spirals in the inclination-corrected sample are shown
in Figure~\ref{fig:qkqb}. Spirals with $q_B \lesssim 0.3$
are generally less flat in the $K_s$ band than in the $B$ band.
It can also been seen that extremely flat spirals,
with $(q_B + q_K)/2 < 0.2$, and nearly round spirals,
with $(q_B+q_K)/2 > 0.9$, are more likely to be late types
(squares in Figure~\ref{fig:qkqb}) than early types
(triangles in Figure~\ref{fig:qkqb}). Intermediate axis
ratios, however, with $(q_B + q_K)/2 \sim 0.5$, are strongly
dominated by early spirals. When galaxies with detectable
visible-light bars (types SAB and SB) are compared to galaxies
of type S and SA, no significant difference is found in
their axis ratio distributions; $P \geq 0.1$ as measured
both by a Kolmogorov-Smirnov test and by a $\chi^2$ test,
in both bands.

\begin{figure}
\plotone{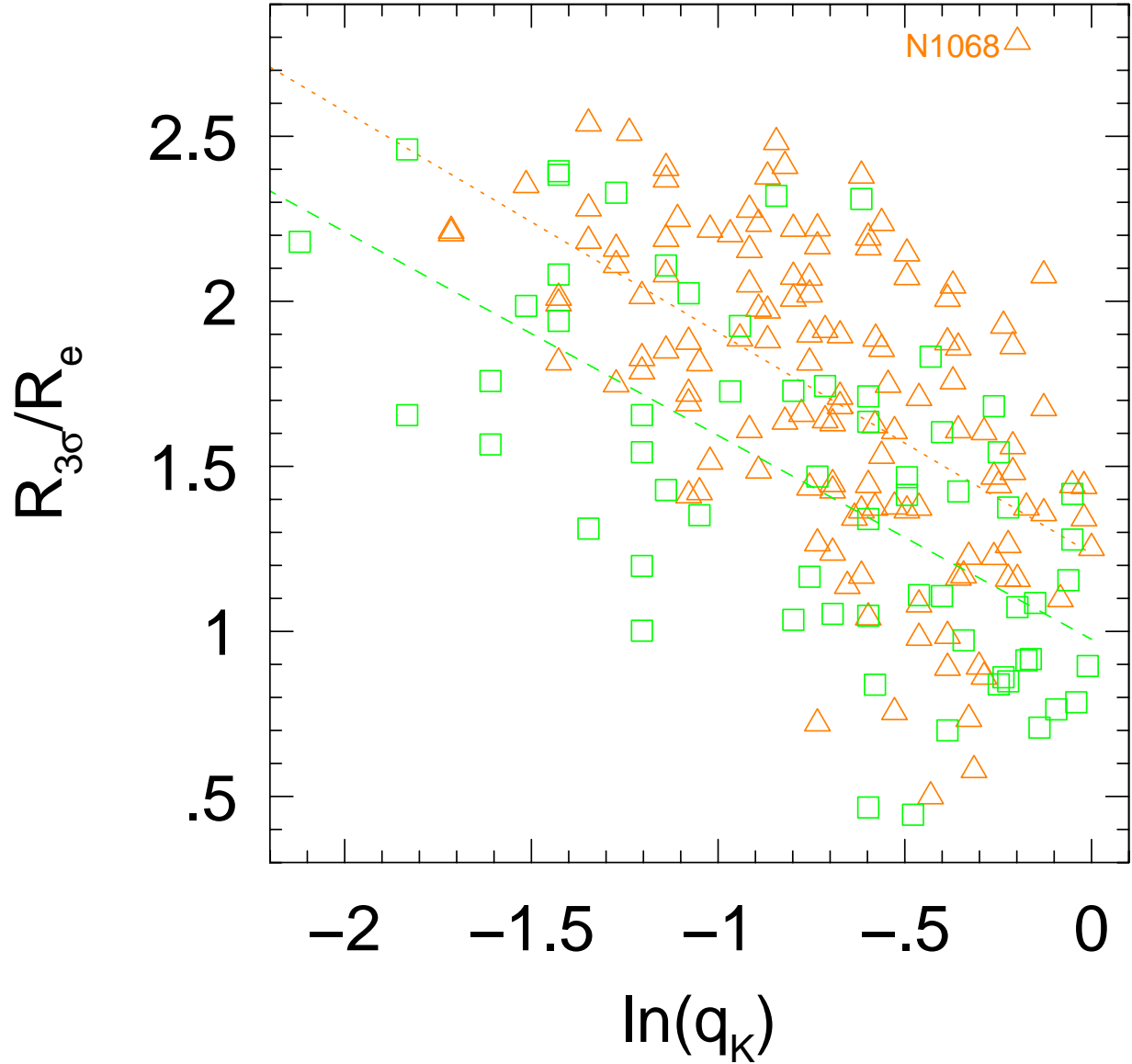}
\caption{The semimajor axis $R_{3\sigma}$ of the $3\sigma$
isophote, measured in units of the half-light radius $R_e$
of the best Sersic profile, plotted
as a function of the axis ratio $q_K$.
Squares indicate late spirals; the dashed line is the best fit
to their $R_{3\sigma}/R_e$ vs. $\ln(q_K)$ relation. Triangles
indicate early spirals; the dotted line is the best fit to their
$R_{3\sigma}/R_e$ vs. $\ln(q_K)$ relation. The labeled
outlier, NGC 1068, is excluded from the fit.
}
\label{fig:r3sig}
\end{figure}

On a cautionary note, it should be pointed out that $q_K$,
the axis ratio of the $3\sigma$ isophote, doesn't measure
the shape at the same physical distance from each galaxy's
center. Because looking at a disk edge-on enhances the
surface brightness, the distance $R_{3\sigma}$, the semimajor
axis of the $3\sigma$ isophote, will be larger when a galaxy
is seen edge-on than when it is seen face-on. Thus, $q_K$
is measuring the shape at larger radii, on average, for
galaxies with small $q_K$ than for galaxies with large $q_K$.
To illustrate the trend, Figure~\ref{fig:r3sig} shows $R_{3\sigma}$,
in units of the effective radius, $R_e = (b_n)^n \alpha$, as
a function of $\ln q_K$. On average,
$R_{3\sigma}/R_e$ tends to increase as $q_K$
decreases. The best linear fit to the data for late spirals,
shown as the dashed line in Figure~\ref{fig:r3sig}, is
\begin{equation}
{R_{3\sigma} \over R_e} = 0.976 - 0.617 \ln q_K \ ,
\end{equation}
with a correlation coefficient $r = -0.64$. The best fit to the data for
early spirals, shown as the dotted line in Figure~\ref{fig:r3sig}, is
\begin{equation}
{R_{3\sigma} \over R_e} = 1.235 - 0.670 \ln q_K \ ,
\end{equation}
with a correlation coefficient $r = -0.56$. Although the
scatter in the relation is large, the
$3\sigma$ isophote tends to be closer to the galaxy's
center for nearly face-on axisymmetric disks than for nearly
edge-on disks. If the isoluminosity surfaces of a disk galaxy
were perfectly similar oblate spheroids, it wouldn't matter
where $q_K$ is measured. However, for early-type, big-bulged spirals,
the value of $q_K$ measured for galaxies of low inclination can
be significantly affected by light from the galaxy's central bulge.
A similar relation holds true between $R_{25}$ and $q_B$, the
apparent axis ratio of the $\mu_B = 25 {\rm\,mag}{\rm\,arcsec}^{-2}$.
However, since $R_{25}$ in the $B$ band is typically
larger than $R_{3\sigma}$ in the $K_s$ band, the value
of $q_B$ is usually not strongly influenced by bulge light.
For the early spirals, the mean and standard deviation
of $R_{25}/R_{3\sigma}$ are $1.90 \pm 0.62$; for the late spirals,
$R_{25}/R_{3\sigma} = 2.45 \pm 1.04$. (For two early spirals at
low galactic latitude, WKK 4748 and Maffei 2, $R_{25} / R_{3\sigma}
\lesssim 0.5$.)

\subsection{Nonparametric fits}

The distributions of apparent axis ratios, $f(q_K)$ and $f(q_B)$,
were estimated for the galaxies in the sample using a nonparametric
kernel technique \citep{vi94,tr95,ry96}. Estimating a function $f$
by the use of kernels gives a smooth estimate function $\hat{f}$ with
no \emph{a priori} restrictions on its shape, unlike a parametric
fit to $f$. Given a sample of $n$ axis ratios,
$q_1$, $q_2$, $\dots$, $q_n$, the kernel estimate of
the frequency distribution $f(q)$ is
\begin{equation}
\hat{f}(q) = {1 \over n} \sum_{i=1}^n {1 \over h_i} K \left(
{q-q_i \over h_i} \right) \ ,
\label{eq:kernel}
\end{equation}
where $K(x)$ is the kernel function, normalized so that its
integral over all $x$ is equal to one. To ensure that the
estimate $\hat{f}$ is smoothly differentiable, I adopt a
Gaussian kernel, $K(x) \propto \exp ( - x^2 / 2)$. The parameter
$h_i$ in equation~(\ref{eq:kernel}) is the kernel width,
determined using the standard adaptive two-stage estimator of
\citet{ab82}. In the two-stage estimator, the first stage
consists of making an initial estimate $\hat{f}_0$ using
a fixed kernel width $h$. The initial kernel width is given
by the formula $h = 0.9 A n^{-0.2}$, with
$A = \min ( \sigma , Q_4 / 1.34)$, where $\sigma$ is the
standard deviation of the axis ratios and $Q_4$ is the interquartile
range. For samples that are not extremely skewed, this formula
minimizes the mean square error of the estimate \citep{si86,vi94}.
The second stage consists of making the final estimate $\hat{f}$,
using at each data point $q_i$ the kernel width
\begin{equation}
h_i = h \left[ {G \over \hat{f}_0 (q_i)} \right]^{1/2} \ ,
\end{equation}
where $G$ is the geometric mean of $\hat{f}_0$ over all values
of $q_i$.

\begin{figure}
\plotone{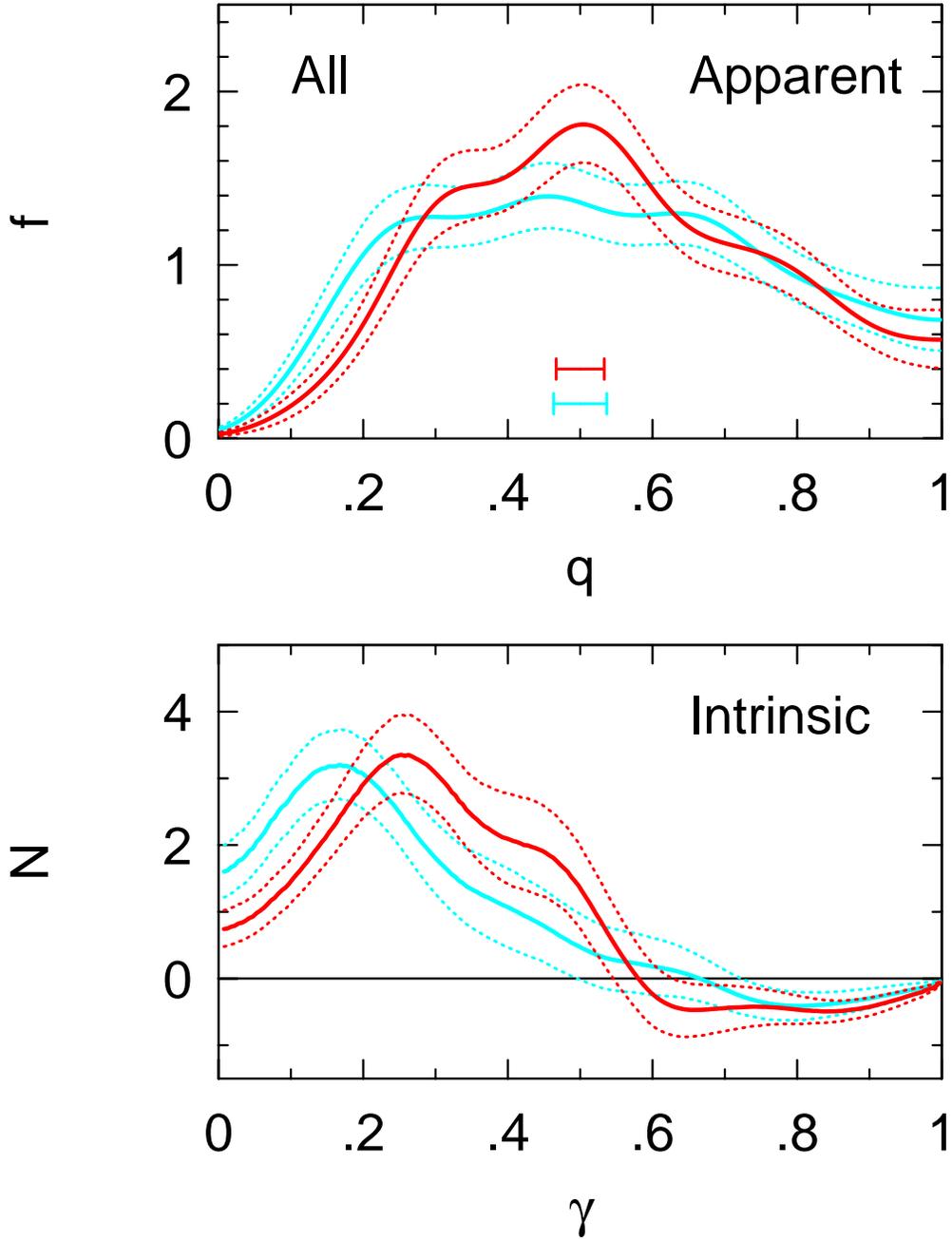}
\caption{\emph{Upper}: Distribution of apparent axis ratios for
the inclination-adjusted sample of 2MASS LGA spirals. Red lines
indicate the $K_s$ axis ratios; blue lines indicate the $B$
axis ratios. The solid line in each case is the best fit, while
the dotted lines indicate the 80\% confidence interval estimated
from bootstrap resampling. The horizontal error bars show the
initial kernel width $h$. \emph{Lower}: Distribution of intrinsic
axis ratios, assuming the spirals are randomly oriented oblate spheroids.
}
\label{fig:all_axi}
\end{figure}

The estimated distribution of axis ratios for the 193 spirals
in the inclination-corrected sample is shown in the upper
panel of Figure~\ref{fig:all_axi}. The solid red line indicates
$\hat{f} (q_K)$, while the solid blue line indicates
$\hat{f} (q_B)$. The dotted lines indicate the 80\% error
intervals estimated by bootstrap resampling. In the bootstrap
analysis, I randomly selected $n = 193$ data points, with substitution,
from the original sample of 193 axis ratios; a new estimate
$\hat{f}$ was then created from the bootstrapped data. After
doing 5000 bootstrap estimates, I determined the 80\% error
interval; in Figure~\ref{fig:all_axi}, 10\% of the bootstrap
estimates lie above the upper dotted line, and 10\% lie below
the lower dotted line. If the spiral galaxies were infinitesimally
thin, perfectly circular disks, the distribution function would
be $f = 1$ for $0 \leq q \leq 1$; the deviations of $f(q_K)$ and
$f(q_B)$ from this form reveal that the spiral galaxies in our
sample are neither infinitesimally thin nor perfectly circular.
In the $K_s$ band, galaxies are slightly rounder on average
than in the $B$ band: $\langle q_K \rangle = 0.538$ versus
$\langle q_B \rangle = 0.525$. In addition, there is a slightly greater
spread in shapes in the $B$ band: $\sigma (q_K) = 0.212$ versus
$\sigma (q_B) = 0.234$.

Given the assumption of oblateness, it is straightforward
to do a numerical inversion of $\hat{f}(q)$, the estimated
distribution of apparent axis ratios, to find $\hat{N} (\gamma)$,
the estimated distribution of intrinsic axis ratios. The
relation between $\hat{f}(q)$ and $\hat{N} (\gamma)$ is
\begin{equation}
\hat{f} (q) = \int_0^q  P ( q | \gamma ) \hat{N} (\gamma) d \gamma \ ,
\label{eq:volt}
\end{equation}
where $P (q | \gamma ) dq$ is the probability that a galaxy with
a short-to-long axis ratio $\gamma$ has an apparent axis ratio
in the range $q \to q + dq$ when seen from a randomly chosen viewing
angle. For a population of oblate spheroids \citep{sa70},
\begin{equation}
P (q | \gamma ) =
{q \over (1 -\gamma^2)^{1/2}(q^2-\gamma^2)^{1/2} }
\end{equation}
if $\gamma \leq q \leq 1$, and $P (q|\gamma) = 0$ otherwise.
Equation~(\ref{eq:volt}) is a Volterra equation of the first kind;
in its discretized form, it can be inverted to find $\hat{N}$ by a 
process of forward substitution (see \citet{vi05} for details).

The bottom panel of Figure~\ref{fig:all_axi} shows the distributions
of intrinsic axis ratios in the $K_s$ band (red line) and the
$B$ band (blue line). The solid line, in each case, is the distribution
found by inverting equation~(\ref{eq:volt}). Because $\hat{N}$ is
a deconvolution of $\hat{f}$, any noise present in $\hat{f}$ will
be amplified in $\hat{N}$; to produce an acceptably smooth estimate
of $N$, it is thus necessary to use a value of the initial kernel
width $h$ that is larger than the optimal value for computing $\hat{f}$
\citep{tr95}. For computing $\hat{N}$, I used a kernel width
$h$ that was 20\% greater than the value used to compute $\hat{f}$.

The deduced distributions $\hat{N}$ show that the most probable value
of $\gamma$ is $\gamma_K = 0.25$ in the $K_s$ band and $\gamma_B = 0.17$
in the $B$ band. A striking aspect of the distributions $\hat{N}$
seen in Figure~\ref{fig:all_axi} is that they are \emph{negative} for
large values of $\gamma$. In the $K_s$ band, $\hat{N}$ dips below zero
for $\gamma_K > 0.58$; in the $B$ band, $\hat{N}$ is below zero for
$\gamma_B > 0.67$. This is a physical absurdity, implying a negative
number of nearly spherical galaxies. The 80\% confidence band, delineated
by the dotted lines in the lower panel of Figure~\ref{fig:all_axi},
falls below zero, in both bands, for $\gamma \geq 0.7$. This means
that fewer than 10\% of the bootstrap resamplings yield non-negative
estimates in this range of $\gamma$. The hypothesis that the
spiral galaxies are a population of randomly oriented oblate
spheroids can thus be rejected at the 90\% (one-sided) confidence
level. If the 98\% confidence band is laid out in a similar manner,
with 1\% of the bootstrap resamplings lying above the band and
1\% lying below, it is found that in the $K_s$ band, the
confidence band dips below zero for large $\gamma$. Thus, in
the $K_s$ band, the oblate hypothesis can be rejected at the
99\% (one-sided) confidence level. In the $B$ band, where
the shape is measured farther from the galaxies' centers, the
oblate hypothesis cannot be rejected at the 99\% confidence level.

\begin{figure}
\plotone{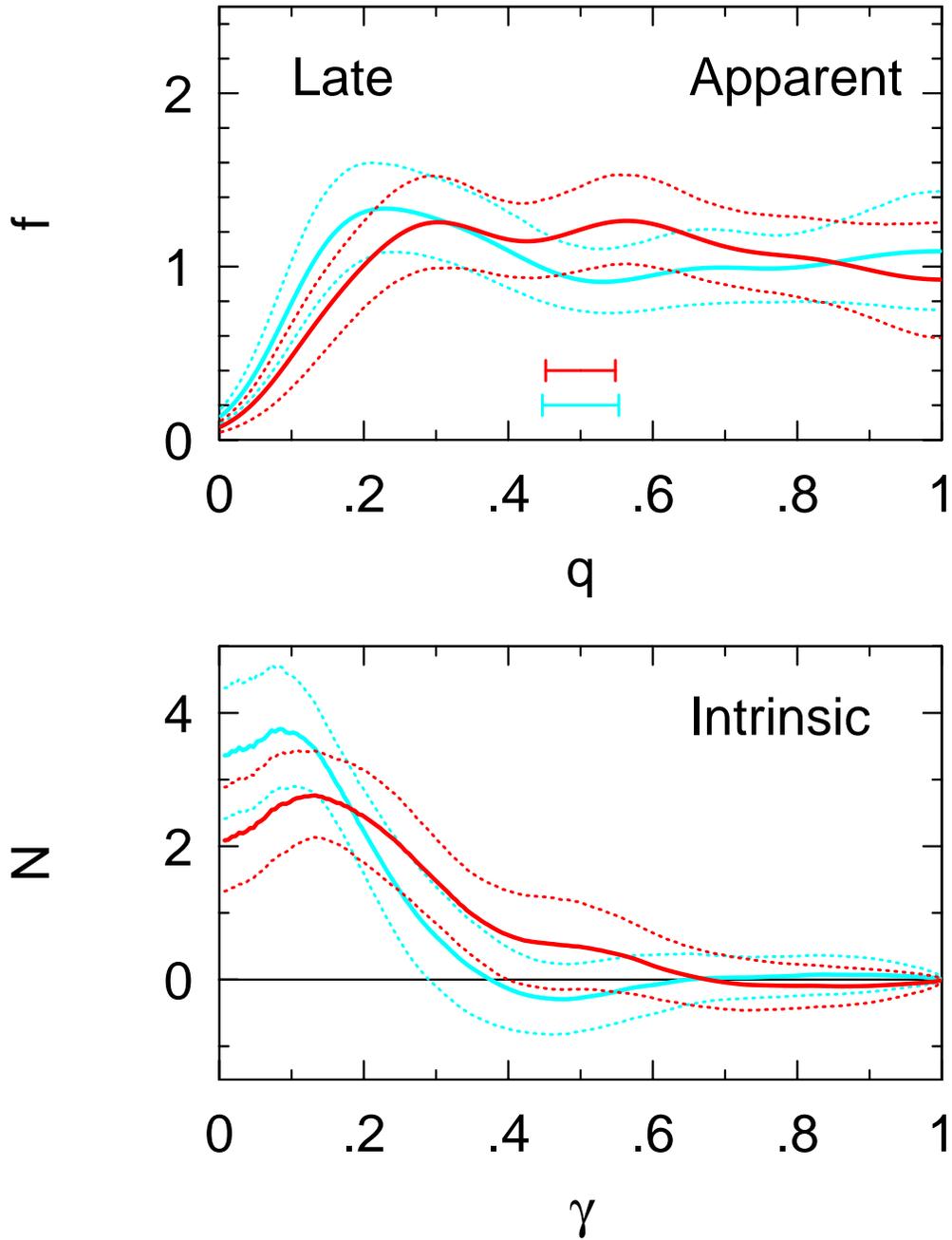}
\caption{The same as Figure~\ref{fig:all_axi}, including only the
$n = 63$ spirals of Hubble type Sc and later.}
\label{fig:late_axi}
\end{figure}

For the complete inclination-adjusted sample, the hypothesis that
the galaxies are perfectly axisymmetric can be rejected at a fairly
strong confidence level, particularly in the near infrared. However,
Figure~\ref{fig:qkqb} reveals that the distribution of
apparent shapes is different for late spirals and for early
spirals. Thus, it is worthwhile to look at $\hat{f} (q)$ and
the derived $\hat{N} (\gamma)$ for late and early spirals separately.
The upper panel of Figure~\ref{fig:late_axi} shows $\hat{f} (q_K)$
and $\hat{f} (q_B)$ for the $n = 63$ late spirals in the sample.
The distribution of apparent axis ratios in the $K_s$ band,
indicated by the red lines, is similar to the distribution
in the $B$ band, indicated by the blue lines. In both cases, the
distribution $\hat{f}$ is nearly constant over a wide range of $q$.
The deduced distributions of intrinsic axis ratios,
$\hat{N} (\gamma_K)$ and $\hat{N} (\gamma_B)$, are
indicated by the red and blue lines respectively in the bottom
panel of Figure~\ref{fig:late_axi}. In both bands,
the distribution $\hat{N}$ never falls below zero; thus, the
oblate hypothesis is perfectly acceptable, both for the $B$ band
data and for the $K_s$ band data. If the oblate hypothesis is
accepted, the average disk thickness is smaller in the $B$ band:
$\langle \gamma_B \rangle = 0.12$ as compared to $\langle \gamma_K
\rangle = 0.19$. In both bands, the width of the function $\hat{N}$
is comparable to the kernel width $h$ used, indicating that
the true spread in $\gamma_B$ and $\gamma_K$ may be less than
is shown in the smoothed distributions of Figure~\ref{fig:late_axi}.

\begin{figure}
\plotone{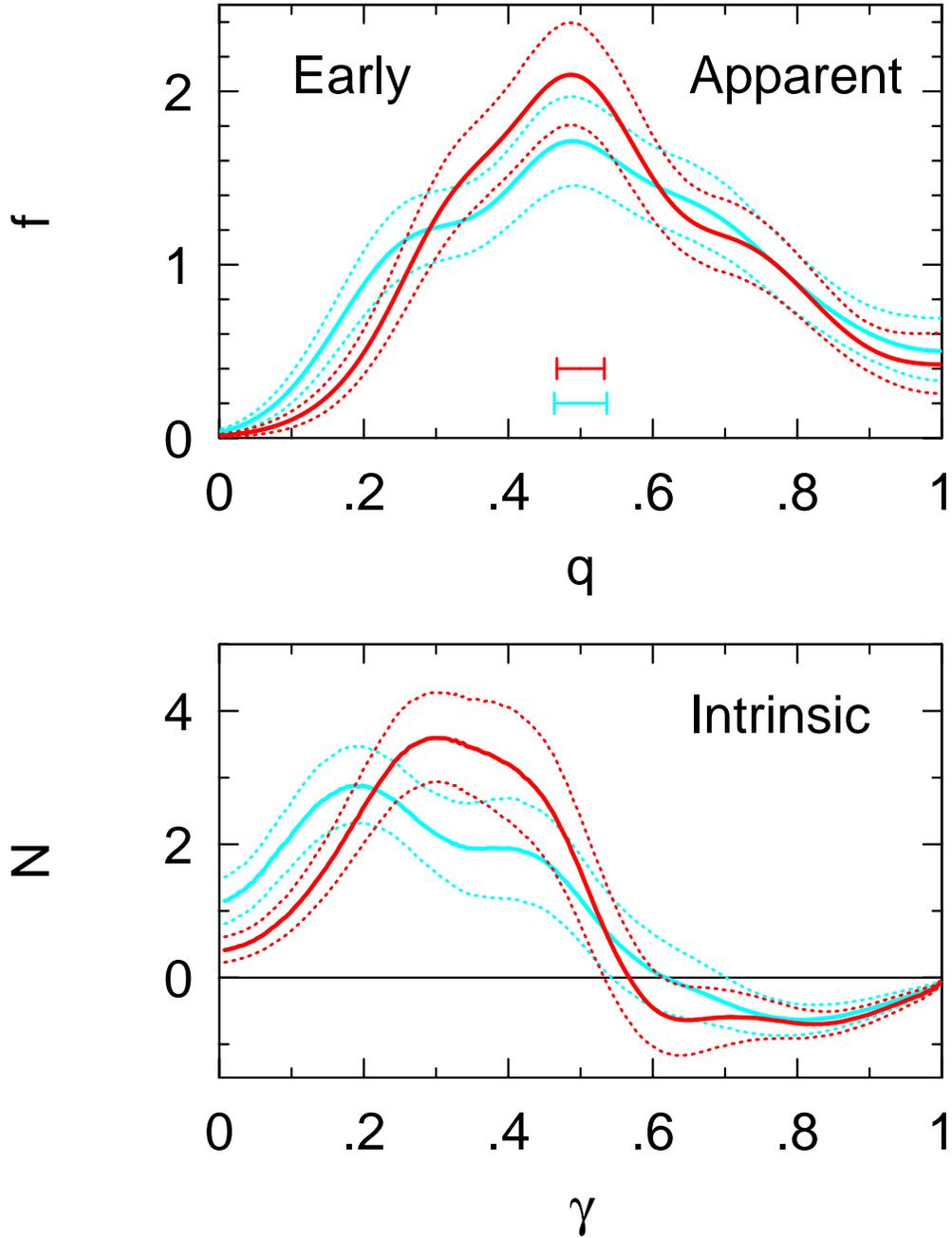}
\caption{The same as Figure~\ref{fig:all_axi}, including only the
$n = 130$ spirals of Hubble type Sbc and earlier.}
\label{fig:early_axi}
\end{figure}

The apparent shapes of early spirals, shown in the upper panel
of Figure~\ref{fig:early_axi}, are strikingly different from the
apparent shapes of late spirals. For early spirals, the distribution
$\hat{f}$ is not uniform over $q$, but instead is peaked at $q \sim 0.5$,
with a marked scarcity of galaxies with $q \gtrsim 0.8$. Such a nonuniform
distribution is typical of a population of triaxial ellipsoids. The
distribution of $q_K$ (red lines in Figure~\ref{fig:early_axi}) is
particularly strongly peaked at $q \sim 0.5$, which may be attributed, in
part, to a contribution of light from nonaxisymmetric bulges. However,
even the distribution of $q_B$, measured at a much larger radius within
each galaxy, is inconsistent with axisymmetry. The deduced distributions
of intrinsic axis ratios, shown in the lower panel of Figure~\ref{fig:early_axi},
demonstrate how grossly inconsistent the observed distributions of $q_K$
and $q_B$ are with the hypothesis of randomly oriented oblate spheroids.
The oblate hypothesis can, in fact, be ruled out at the 99\% confidence
level for both the $K_s$ band isophotal shapes and the $B$ band isophotal
shapes. Thus, the deviations from axisymmetry in the full sample of
spiral galaxies can be attributed to the early spirals in the sample;
the late spirals are perfectly consistent with axisymmetry.

\subsection{Parametric fits}

For comparison with previous studies, it is useful to make parametric
estimates of $N (\gamma)$, in addition to the nonparametric kernel
estimates of the previous subsection. A common convention \citep{hu92,fa93}
is to fit the distribution of intrinsic axis ratios with a Gaussian:
\begin{equation}
N (\gamma)  d\gamma \propto \exp \left[ - {(\gamma - \mu_\gamma)^2
\over 2 \sigma_\gamma^2} \right] d\gamma
\label{eq:ngam}
\end{equation}
for $0 \leq \gamma \leq 1$, with $N = 0$ elsewhere. Inspection of the
lower panel of Figure~\ref{fig:late_axi} indicates that this should
provide a reasonable fit for the late spirals, at least.

The best-fitting values of $\mu_\gamma$ and $\sigma_\gamma$ were
determined by a $\chi^2$ fit to the binned distribution of $q_K$
and $q_B$. In each band, the $n$ values of $q$ were divided into
$n_{\rm bin} = 12$ bins of equal width. After choosing particular
values for $\mu_\gamma$ and $\sigma_\gamma$, I randomly chose
$n$ values of $\gamma$, drawn from the distribution of
equation~(\ref{eq:ngam}). After randomly selecting a viewing angle
for each mock disk, I computed its apparent axis ratio
$q (\gamma,\theta,\phi)$ from the formulas in \S\ref{sec-sample}.
The model axis ratios were then binned in the same way as the observed
axis ratios. After repeating this procedure 8000 times for
a given $(\mu_\gamma,\sigma_\gamma)$ pair, I calculated the mean
and standard deviation of the expected number of model galaxies
in each bin, and computed a $\chi^2$ probability for that
particular pair of parameters.

\begin{figure}
\plotone{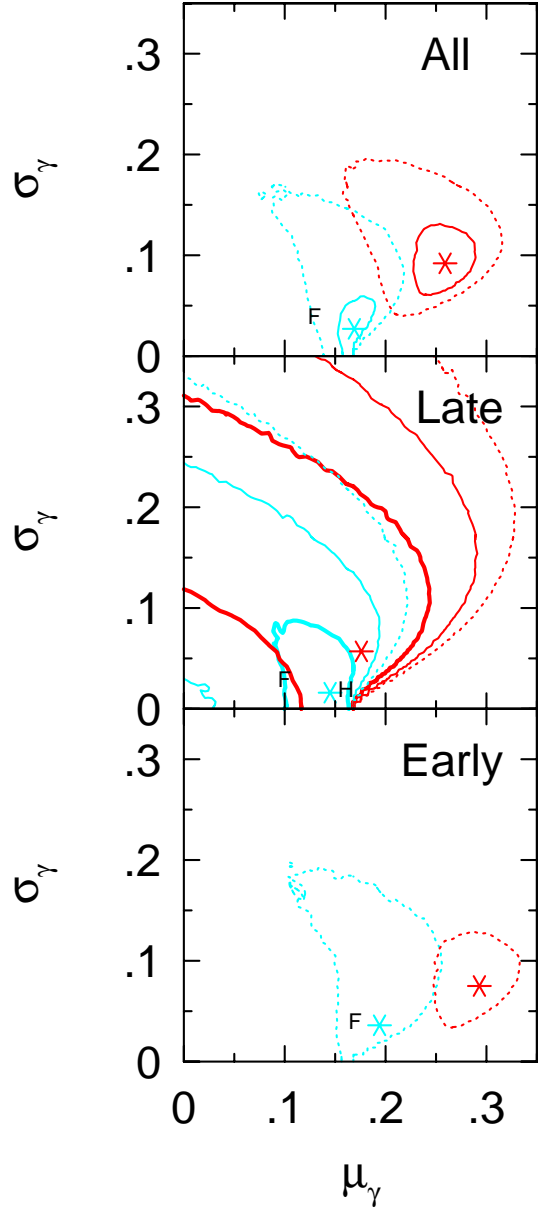}
\caption{\emph{Top}: Asterisks indicate the best fitting
values of $\mu_\gamma$ and $\sigma_\gamma$ to the complete
sample of LGA spirals; the heavy solid line, light
solid line, and dotted line indicate the $P = 0.5$, $P = 0.1$,
and $P = 0.01$ isoprobability contours, found by a
$\chi^2$ fit. Red = shapes deduced from
$K_s$ data; blue = shapes deduced from $B$ data. \emph{Middle}:
Same, but for spirals of Hubble type Sc or later.
\emph{Bottom}: Same, but for spirals of type Sbc or earlier.
`F' = fits of \citet{fa93}. `H' = fit of \citet{hu92}
for Sc spirals.
}
\label{fig:gaus_axi}
\end{figure}

Figure~\ref{fig:gaus_axi} shows the isoprobability contours in
the two-dimensional parameter space, where the probability is
measured by a $\chi^2$ fit to the binned data. The fit to the
$K_s$ band data is shown by the red lines; the fit to the
$B$ band data is shown by the blue lines. In each band, the best
fitting values of $\mu_\gamma$ and $\sigma_\gamma$ are indicated
by an asterisk.

In the top panel of Figure~\ref{fig:gaus_axi}, the fit
is made to the complete inclination-adjusted sample of $n = 193$
spirals. In the $K_s$ band, the best fit in this region of parameter
space is $(\mu_\gamma,\sigma_\gamma) = (0.26,0.09)$, which yields
a $\chi^2$ probability $P \approx 0.20$. In the $B$ band, the
best fit is $(\mu_\gamma,\sigma_\gamma) = (0.17,0.03)$, which yields
$P \approx 0.17$. For these fits, the biggest contributions to
$\chi^2$ comes from the two bins with $q \geq 0.83$, for which
this axisymmetric model predicts more galaxies than are observed.
In the $K_s$ band, spiral galaxies are not only thicker, on average,
than in the $B$ band, but they also have a greater spread in thicknesses.
The typical $\gamma$ I find for spirals in the $B$ band,
$\mu_\gamma = 0.17$, is greater than the value of $\mu_\gamma = 0.13$
found by \citet{fa93} for a volume-limited sample of 766 spirals,
using apparent axis ratios taken from the Third Reference Catalogue
of Bright Galaxies \citep{de91}. As seen in the upper panel of
Figure~\ref{fig:gaus_axi}, the best values of $(\mu_\gamma,\sigma_\gamma)$
found by \citet{fa93} lie just inside the $P = 0.01$ contour for
the 2MASS LGA spirals.

The middle panel of Figure~\ref{fig:gaus_axi} shows the fit
for the late spirals only (Hubble type Sc and later). The $n = 63$
galaxies in this subsample do not place a strong constraint on
$\mu_\gamma$ and $\sigma_\gamma$; the $P = 0.5$ contour, indicated
by the heavy solid line, contains a large swath of parameter space,
particularly in the $K_s$ band. The best fit in the $K_s$
band is $(\mu_\gamma,\sigma_\gamma) = (0.18,0.06)$, which yields
a $\chi^2$ probability $P \approx 0.91$. In the $B$ band, the best
fit is $(\mu_\gamma,\sigma_\gamma) = (0.14,0.02)$, yielding
$P \approx 0.80$. Once again, as for the complete sample, the
$K_s$ band data indicates both a larger average thickness and a
greater spread in thicknesses than does the $B$ data. However,
in both bands, the data are consistent with $\sigma_\gamma = 0$.
An excellent fit results if all the late spirals are assumed
to have an intrinsic axis ratio $\gamma$ equal to, or slightly
smaller than, the smallest apparent axis ratio $q$ in the subsample
($q_K = 0.12$ for NGC 4302, and $q_B = 0.12$ for NGC 5907).
The best fit found by \citet{hu92} for a sample of Sc galaxies,
$(\mu_\gamma,\sigma_\gamma) = (0.16,0.02)$, and by \citet{fa93}
for a sample of galaxies of type $4.5 < t \leq 7.0$,
$(\mu_\gamma,\sigma_\gamma) = (0.10,0.03)$, are both
statistically consistent with the results for late-type
LGA 2MASS spirals in the $B$ band; see the middle panel of
Figure~\ref{fig:gaus_axi}.

The bottom panel of Figure~\ref{fig:gaus_axi} shows the results
of fitting a Gaussian distribution for $\gamma$ to the
early spirals (Hubble type Sbc and earlier). The best
fits are not as good as the excellent fits found
for the late spirals. The best fit in the $K_s$
band is $(\mu_\gamma,\sigma_\gamma) = (0.29,0.08)$, for
which the $\chi^2$ probability is $P \approx 0.06$. The
best fit in the $B$ band is $(\mu_\gamma,\sigma_\gamma) =
(0.20,0.04)$, yielding $P \approx 0.08$. The relatively poor
fits are the result of demanding axisymmetry; the Gaussian form
of equation~(\ref{eq:ngam}) is not primarily to blame, since the
nonparametric kernel estimate revealed that no distribution
$N (\gamma) > 0$ gives a good fit to the early spirals.
As with the late spirals, the average thickness and the
spread of thicknesses is greater in the $K_s$ band than
in the $B$ band. The greater disk thickness in the infrared
is more pronounced for the early spirals than for the late
spirals. For early spirals, the best-fitting $\mu_\gamma$ in
the $K_s$ band is 50\% greater than in the $B$ band; for late
spirals, the best-fitting $\mu_\gamma$ in the $K_s$ band is
only 20\% greater than in the $B$ band.

\section{Shape of Nonaxisymmetric Disks}
\label{sec-nonaxi}

Given that a distribution of axisymmetric ($\epsilon = 0$) disks
does not give a good fit to early spirals, it would be useful to
know which distribution of nonaxisymmetric disks gives the best
fit to the early spirals (and to the late spirals, for that matter).
A fundamental difficulty is that even if $f(q)$ were known
perfectly, it would not uniquely determine $N(\gamma,\epsilon)$.
However, the joint distribution of the disk thickness $\gamma$ and
the disk ellipticity $\epsilon$ is not beyond all conjecture. I will
use the approach of assuming a parameterized functional form for the
distribution $N(\gamma,\epsilon)$, and finding which values of the
parameters give the best fit to the observed distributions of axis ratios.
This approach yields permissible fits to $N(\gamma,\epsilon)$, but
these fits, it should be remembered, are not unique solutions.

\subsection{Gaussian parametric fits}

I will assume that $N(\gamma,\epsilon) = N_\gamma (\gamma) \times
N_\epsilon (\epsilon)$, where $N_\gamma$ is the same Gaussian
assumed in equation~(\ref{eq:ngam}) and $N_\epsilon$ is
a Gaussian distribution \citep{la92,hu92,fa93,ry04}:
\begin{equation}
N_\epsilon (\epsilon) d\epsilon \propto \exp \left[ - {(\epsilon - \mu_\epsilon)^2
\over 2 \sigma_\epsilon^2} \right]  d\epsilon \ .
\label{eq:gau}
\end{equation}
The function $N(\gamma,\epsilon)$ is nonzero inside the
region $0 \leq \gamma \leq 1$ and $0 \leq \epsilon \leq 1-\gamma$.
A complication of fitting nonaxisymmetric models is the
increased difficulty of deriving an inclination-corrected
subset of spiral galaxies. If a galaxy is not an axisymmetric
disk, it is impossible from photometry alone to 
determine absolutely whether or not its face-on semimajor
axis $R_{20}^o$ is greater than the cutoff value of $60\arcsec$.
To illustrate the effects of nonaxisymmetry,
consider a mildly triaxial disk, with
$\epsilon \ll 1$. For such a disk, $\beta = 1-\epsilon$,
$D \approx 1 + 2 \epsilon \cos^2 \phi$, and the face-on
semimajor axis is, from equation~(\ref{eq:incor}),
\begin{equation}
R_{20}^o \approx ( 1 + \epsilon \cos^2 \phi )
R_{20} \left[ 1 + ( \alpha / R_{20} )^{1/n} ( \ln q - \epsilon \cos 2 \phi )
\right]^n \ .
\label{eq:neps}
\end{equation}
Thus, the face-on isophotal radius $R_{20}^o$ of a triaxial disk
may be either smaller or larger than that of an axisymmetric disk
with the same observed values of $R_{20}$, $\alpha$, $n$, and $q$.
Since $R_{20}^o$ can be greater than the observed $R_{20}$ for a
nonaxisymmetric disk, a sample complete to $R_{20} = 60 \arcsec$ is
not guaranteed to be complete to $R_{20}^o = 60 \arcsec$. However, for
mildly eccentric disks ($\epsilon \ll 1$), the potential incompleteness
effects should be small.

To make the inclination correction, I begin by assuming values
for the parameters $\mu_\gamma$, $\sigma_\gamma$, $\mu_\epsilon$,
and $\sigma_\epsilon$. I randomly select a value of $\gamma$ and of
$\epsilon$ from the appropriate distribution functions, and
randomly choose a viewing angle $(\theta,\phi)$. I then compute
$\beta = 1 - \epsilon$, $D(\theta,\phi)$, using
equation~(\ref{eq:d}), $E(\theta,\phi)$, using equation~(\ref{eq:e}),
and the apparent axis ratio $q = \sqrt{D/E}$. I then
have one possible $(\beta,D)$ pair corresponding to the observable
apparent axis ratio $q$. Having binned the entire range of $q$, from
$q = 0$ to $q = 1$, into 100 equal width bins, I can then place the
$(\beta,D)$ pair into the appropriate bin. I repeat the process of
randomly choosing $\gamma$, $\epsilon$, $\theta$, and $\phi$,
computing $\beta$, $D$, $E$, and $q$, and putting the
resulting $(\beta,D)$ pair into the appropriate bin in $q$, until
I have done a total of $\sim 5 \times 10^6$ iterations of this process,
keeping a maximum of $1000$ $(\beta,D)$ pairs for each bin in $q$.
Now consider a specific galaxy with measured values of $R_{20}$,
$\alpha$, $n$, and $q_K$. I go to the appropriate bin in axis ratio $q$
and pull out $(\beta,D)$ pairs, determining how many of the pairs
result in $R_{20}^o > 60 \arcsec$, using equation~(\ref{eq:incor}).
A fraction $F$ of the $(\beta,D)$ pairs, where $0 \leq F \leq 1$, will
result in the face-on isophotal radius $R_{20}^o$ of this galaxy being greater
than the lower limit of $60 \arcsec$. I then created binned distributions
of $q_K$ and $q_B$, with $n_{\rm bin} = 12$, for comparison with
model distributions. In the binned distributions of observed
axis ratios, the axis ratio for each galaxy is given a weight equal
to the fraction $F$. As a consequence, the total number $n = \sum F$ of
galaxies in the binned distribution may be non-integral. For convenience,
I round the number $n$ to the nearest integer by multiplying the
number of galaxies per bin by the factor $C = {\rm\,int}(n+0.5)/n$.

Finding the global best fit in the four-dimensional parameter
space $(\mu_\gamma,\sigma_\gamma,\mu_\epsilon,\sigma_\epsilon)$
involves a tedious search. The search can be simplified by
noting that for relatively thin, mildly elliptical disks, the
values of $\mu_\gamma$ and $\sigma_\gamma$ are determined by
looking at galaxies that are close to edge-on, with $q < 0.5$;
by contrast, the values of $\mu_\epsilon$ and $\sigma_\epsilon$
are fixed by looking at galaxies that are close to face-on,
with $q > 0.5$. Thus, the best-fitting values of $\mu_\gamma$
and $\sigma_\gamma$ that I found by assuming $\epsilon = 0$
should not be greatly changed when $0 < \epsilon \ll 1$. I
therefore fix the values of $\mu_\gamma$ and $\sigma_\gamma$
for each sample of galaxies to be equal to those determined
in the previous subsection, where galaxies were assumed to
be axisymmetric. This reduces the parameter space to
two dimensions. After fixing $\mu_\gamma$ and $\sigma_\gamma$,
I choose values of $\mu_\epsilon$ and $\sigma_\epsilon$. I
then randomly select $n$ values of $\gamma$ from the
distribution of equation~(\ref{eq:ngam}), $n$ values of
$\epsilon$ from the distribution of equation~(\ref{eq:neps}),
and $n$ random viewing angles. The apparent axis ratio
$q (\gamma,\epsilon,\theta,\phi)$ for each of $n$ mock galaxies
is computed from the formulas given in \S\ref{sec-sample}.
The model axis ratios are then binned into $n_{\rm bin} = 12$
equal-width bins. After repeating this procedure $\sim 10^4$ times
for a given $(\mu_\epsilon,\sigma_\epsilon)$ pair, I calculate
the mean and standard deviation of the expected number of mock
galaxies in each bin, and compute a $\chi^2$ probability for
that particular set of parameters.

Figure~\ref{fig:gaus_non} shows the isoprobability contours in
$(\mu_\epsilon,\sigma_\epsilon)$ parameter space. The fit to the
$K_s$ band data is shown by the red lines; the fit to the $B$
band data is shown by the blue lines. In each band, the best fitting
values of $\mu_\epsilon$ and $\sigma_\epsilon$ in the displayed
region of parameter space are indicated by an asterisk.
The best fitting $\mu_\epsilon$ and $\sigma_\epsilon$ for each
galaxy sample, at each wavelength, are given in Table~\ref{tab:param}.
The upper panel of Figure~\ref{fig:gaus_non} displays the results
for the complete sample of spirals, weighted by the appropriate
fraction $F$ for each galaxy. The best fits have $\mu_\epsilon
\approx 0$, and $\sigma_\epsilon = 0.16$ ($B$ band) and $\sigma_\epsilon
= 0.26$ ($K_s$ band). The most probable ellipticity for these
skewed distributions is $\epsilon \approx 0$, but there is a long
tail to higher ellipticities. The results for the 2MASS LGA spirals
are in good agreement with those of \citet{la92}, \citet{fa93}, and
\citet{ry04}, who find good fits for Gaussian distributions with
$\mu_\epsilon = 0$, and $\sigma_\epsilon \sim 0.2$.
(I only computed solutions with $\mu_\epsilon \geq 0$. It is
mathematically permissible to have $\mu_\epsilon < 0$. In the
limit $- \mu_\epsilon \gg \sigma_\epsilon$, this leads to an exponential
distribution function: $N_\epsilon \propto \exp (- \epsilon / h_\epsilon )$,
with a scale length
$h_\epsilon = \sigma_\epsilon^2 / |\mu_\epsilon| \ll \sigma_\epsilon$.
Exponential functions peaking at $\epsilon = 0$ do not give a significantly
better fit than Gaussian functions peaking at $\epsilon = 0$.)

The middle panel of Figure~\ref{fig:gaus_non} shows the results of
fitting a Gaussian distribution of ellipticities to the late spirals
(Sc and later). This plot emphasizes that the apparent shapes of late
spirals are consistent with axisymmetry. The best fit in each
band has $\mu_\epsilon \leq 0.07$ and $\sigma_\epsilon = 0.02$,
but $\mu_\epsilon = 0$, $\sigma_\epsilon = 0$ also gives an excellent
fit, with $P > 0.5$. Although perfect axisymmetry is not compulsory, there
are constraints on how large the disk ellipticities can be. The region in
parameter space for which $P > 0.5$ (demarcated by the heavy solid lines in
the middle panel of Figure~\ref{fig:gaus_non}) is confined within the limits
$\mu_\epsilon \leq 0.10$, $\sigma_\epsilon \leq 0.15$. The parameters
found by \citet{fa93} for their sample of late-type galaxies, $\mu_\epsilon = 0$
and $\sigma_\epsilon = 0.09$, are in good agreement with our results.
However, the parameters found by \citet{hu92} for their sample of Sc
galaxies, $\mu_\epsilon = 0.12$ and $\sigma_\epsilon = 0.11$, imply
a greater mean ellipticity than is found in the 2MASS LGA sample.

The bottom panel of Figure~\ref{fig:gaus_non} shows the results of
fitting a Gaussian distribution of ellipticities to the early spirals
(Sbc and earlier). The best-fitting values of $\sigma_\epsilon$ indicate
a broad range of thicknesses: $\sigma_\epsilon = 0.22$ in the $B$ band,
and $\sigma_\epsilon = 0.47$ in the $K_s$ band. Although late spirals
have excellent fits with $\sigma_\epsilon = 0$, getting a good fit
to the observed shapes of early spirals requires $\sigma_\epsilon > 0$.
With a broad tail stretching toward $\epsilon = 1 - \gamma$ (indicating
a prolate shape), the best fit for early spirals in the $K_s$ band
doesn't really represent a population of disks at all. If I quantify
the triaxiality by the parameter $T \equiv (1-\beta^2)/(1-\gamma^2)$,
the best fit in the $K_s$ band, $(\mu_\gamma,\sigma_\gamma,\mu_\epsilon,
\sigma_\epsilon) = (0.29,0.08,0.02,0.27)$, produces a mean triaxiality
parameter $\langle T \rangle = 0.51$, closer to a prolate shape ($T = 1$)
than an oblate shape ($T = 0$). The best fit to early spirals in the
$B$ band also has a significant mean triaxiality, $\langle T \rangle =
0.37$. For comparison, the mean triaxiality for the best fit to
late spirals is $\langle T \rangle = 0.04$ in the $K_s$ band and
$\langle T \rangle = 0.13$ in the $B$ band.

\subsection{Lognormal parametric fits}

Satisfactory fits to the distribution of apparent axis ratios can
also be found using a lognormal distribution of ellipticities, rather
than a Gaussian. If I define $\eta \equiv \ln \epsilon$,
then a lognormal distribution can be written in the form
\begin{eqnarray}
N_\epsilon (\epsilon) d\epsilon & \propto & \exp \left( -
{(\eta - \mu_\eta)^2 \over 2 \sigma_\eta^2} \right) d\eta \nonumber \\
& \propto & \exp \left(-{(\ln\epsilon-\mu_\eta)^2 \over 2\sigma_\eta^2}
\right) {d\epsilon\over\epsilon} \ .
\label{eq:logn}
\end{eqnarray}
Motivation for the lognormal fit is provided by \citet{an01} and
\citet{an02}, who find, by combining photometric and kinematic
information for nearly face-on spirals, that a lognormal distribution
provides a good fit to the observed disk ellipticities. Finding the
best values of $\mu_\eta$ and $\sigma_\eta$ for each sample of spirals
is done using the technique outlined in the previous subsection. Once
again, the values of $\mu_\gamma$ and $\sigma_\gamma$ used were those
that gave the best fit in the axisymmetric case.

Figure~\ref{fig:lognorm_non} displays the isoprobability contours in
$(\mu_\eta,\sigma_\eta)$ space. For all subsamples and both colors,
the lognormal distribution of ellipticities (equation~\ref{eq:logn}) can
provide as good a fit as the Gaussian distribution (equation~\ref{eq:gau}). The
best-fitting parameters $\mu_\eta$ and $\sigma_\eta$, as measured by a $\chi^2$
fit to the binned data, are indicated by asterisks in Figure~\ref{fig:lognorm_non}
and are listed in Table~\ref{tab:param}. The best lognormal fit to each
subsample yields a median ellipticity that is within $\sim 15\%$ of that provided
by the best Gaussian fit. For all spirals, the median ellipticity is $\epsilon \approx
0.10$ in the $B$ band and $\epsilon \approx 0.16$ in the $K_s$ band.

The top panel of Figure~\ref{fig:lognorm_non} shows the results for the
complete sample of spirals. The best fit in the $B$ band, $\mu_\eta = -2.36$
and $\sigma_\eta = 0.96$, is seen to be similar to the best fit found
by \citet{ry04} for a sample of exponential galaxies in the Sloan Digital
Sky Survey ($\mu_\eta = -2.06$ and $\sigma_\eta = 0.83$ in the $i$ band).
It is also in agreement with the best fit to the sample of \citet{an02},
corrected for their selection criteria ($\mu_\eta = -2.29$ and $\sigma_\eta
= 1.04$). The best fit in the $K_s$ band, $\mu_\eta = -0.14$ and
$\sigma_\eta = 2.74$, requires a much broader spread in ellipticities in
addition to a greater median ellipticity. However, the values of $\mu_\eta$
and $\sigma_\eta$ are not very strongly constrained; excellent fits, with
$P > 0.5$, are produced over a wide range of parameter space with $\mu_\eta
\gtrsim -1$ and $\sigma_\eta \gtrsim 1$.

The middle panel of Figure~\ref{fig:lognorm_non} shows the results for
the late spirals. The best-fitting lognormal distribution in each band
implies a small amount of disk ellipticity; the median ellipticity of
late spirals is $\epsilon \approx 0.07$ in the $B$ band and $\epsilon
\approx 0.02$ in the $K$ band. However, the shapes of late spirals are
seen, once again, to be consistent with axisymmetry ($\mu_\eta \to \infty$,
$\sigma_\eta = 0$). The only region of parameter space that is strongly
excluded is the lower right corner of the panel, where $\mu_\eta$ is
large and $\sigma_\eta$ is small. Finally, the bottom panel of
Figure~\ref{fig:lognorm_non} shows the results for the early spirals.
The best fit in the $B$ band implies a median ellipticity of $\epsilon
\approx 0.18$ for the early spirals. The best fit in the $K_s$ band
implies a still greater degree of disk ellipticity, with a median
ellipticity of $\epsilon \approx 0.30$.

\section{Implications and Discussion}
\label{sec-disc}

The 2MASS Large Galaxy Atlas provides a sample of nearby spiral galaxies with
large angular size ($R_{3\sigma} > 60 \arcsec$ at $\lambda \sim 2$ microns).
In this paper, I have used this relatively small sample of very well-resolved
galaxies to estimate the distribution of apparent and intrinsic shapes of spiral
galaxies. An intriguing result of this study is the difference in disk
ellipticity $\epsilon$ between disk galaxies of early and late Hubble type.

If the entire population of 2MASS LGA spirals is considered, summed
over all Hubble types, the distribution of ellipticities in the $B$ band
is consistent with earlier photometric studies using larger sample sizes
\citep{la92,fa93,ry04}. The $K_s$ band ellipticity is greater, but the shape of
the $3\sigma$ isophote in the $K_s$ band can be influenced by light from triaxial
bulges. The distribution $N_\epsilon (\epsilon)$ for shapes in the $B$ band can
be modeled as a Gaussian peaking at $\mu_\epsilon \approx 0$ and width
$\sigma_\epsilon \approx 0.16$ or as a lognormal distribution with
$\mu_\eta \approx -2.4$ and $\sigma_\eta \approx 1.0$; both these
distributions imply a mean ellipticity $\epsilon \approx 0.1$. When the
late spirals, consisting mainly of Hubble type Sc, are examined, they are
found to be perfectly consistent with axisymmetry in both the $B$
and the $K_s$ band. The early spirals, consisting mainly of types Sb
and Sbc, show a high degree of ellipticity, even in the $B$ band, in
which the mean ellipticity is $\epsilon \approx 0.18$.

It has been noted by \citet{fr92} that a disk ellipticity of
$\epsilon > 0.1$ cannot be reconciled with the observed
small scatter in the Tully-Fisher relation \citep{tu77} if
the disk ellipticity is assumed to trace the potential ellipticity.
Suppose that stars and gas are on closed orbits in a logarithmic
potential with rotation speed $v_c$. If the potential has
a small ellipticity $\epsilon_\phi \ll 1$ in the orbital
plane, then the integrated line profile from the disk will
have a width $W = 2 v_c (1 - \epsilon_\phi \cos 2 \phi ) \sin \theta$
when viewed from a position angle $(\theta,\phi)$ \citep{fr92}.
The difference in line width from that produced in a purely
circular disk will produce a scatter in the observed Tully-Fisher
relation. If all disks have $\epsilon_\phi = 0.1$, the expected scatter
is 0.3 mag, even if the inclination has been determined accurately
from kinematic information \citep{fr92}. In most studies of the
Tully-Fisher relation, the inclination is determined photometrically,
from the apparent axis ratio of the disk, assuming (perhaps
erroneously) that the disk is axisymmetric. If inclinations
are determined in this way, the scatter in the Tully-Fisher
relation will be even greater. 

If galaxies had infinitesimally thin circular disks, the
relation between apparent axis ratio $q$ and inclination
$\theta$ would be $q = \cos \theta$. If the thin disks
actually have ellipticity $\epsilon$ and are viewed at high
inclination ($\sin^2 \theta \gg 2 \epsilon$), the apparent
axis ratio will be $q \approx \cos \theta ( 1 - \epsilon
\cos 2 \phi )$. Thus, if the inclination is estimated by
the relation $\theta_{\rm phot} \equiv \cos^{-1} q$, a
fractional error $\propto \epsilon \cos 2 \phi$ will
be introduced into $\cos \theta_{\rm phot}$ and an error
$\propto \epsilon \cot^2 \theta \cos 2 \phi$ will be
introduced into $\sin \theta_{\rm phot}$. Since the observed
velocity width $W$ must be divided by $\sin \theta$ to find the rotation
speed $v_c$, the error produced by using $\sin \theta_{\rm phot}$
instead of $\sin \theta$ will be negligible only when
$\cot^2 \theta \ll 1$. Thus, most studies of the
Tully-Fisher relation use galaxies of high inclination
($\theta_{\rm phot} \gtrsim 45^\circ$). The Tully-Fisher relation
for nearly face-on galaxies (see, for instance, \citet{an03}) can
only be determined if the inclinations are determined kinematically.

\begin{figure}
\plotone{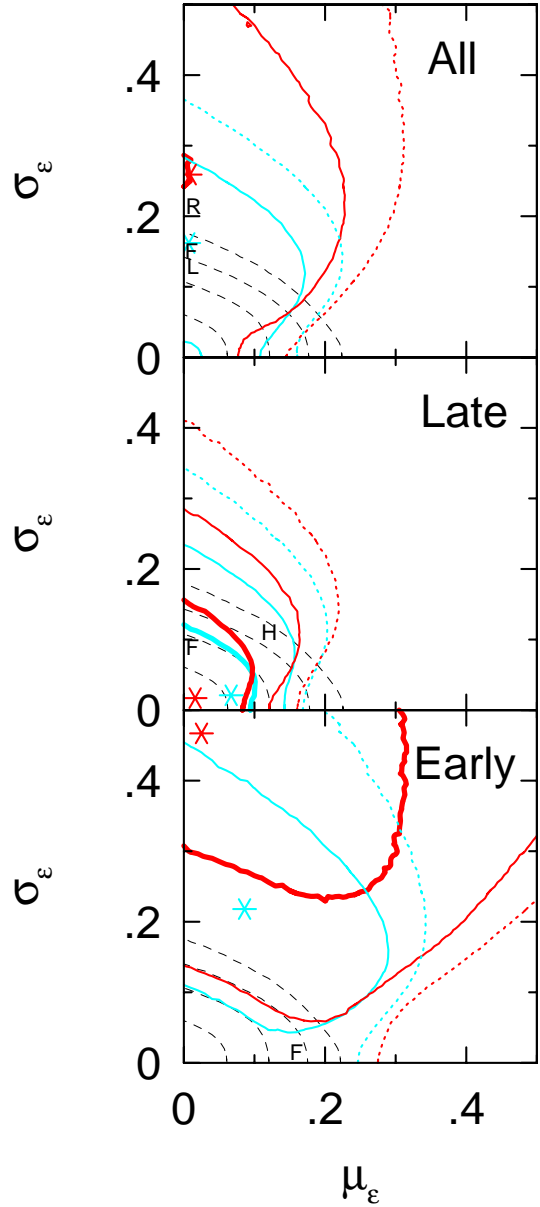}
\caption{\emph{Top}: Asterisks indicate the best fitting values
of $\mu_\epsilon$ and $\sigma_\epsilon$ to the complete sample
of LGA spirals.
The heavy solid line, light solid line, and dotted line
indicate the $P = 0.5$, $P = 0.1$, and $P = 0.01$
isoprobability contours. Red = shapes from
$K_s$ data; blue = shapes from $B$ data.
Dashed lines show scatter in the Tully-Fisher relation
(see \S\ref{sec-disc}); contours are drawn, starting
at the lower left, at 0.25, 0.5, 0.75, and 1.0 magnitude.
\emph{Middle}: Same, but for spirals of Hubble type Sc or later.
\emph{Bottom}: Same, but for spirals of type Sbc or earlier.
`F' = \citet{fa93}; `L' = \citet{la92};
`R' = \citet{ry04}; `H' = \citet{hu92}.
}
\label{fig:gaus_non}
\end{figure}

\begin{figure}
\plotone{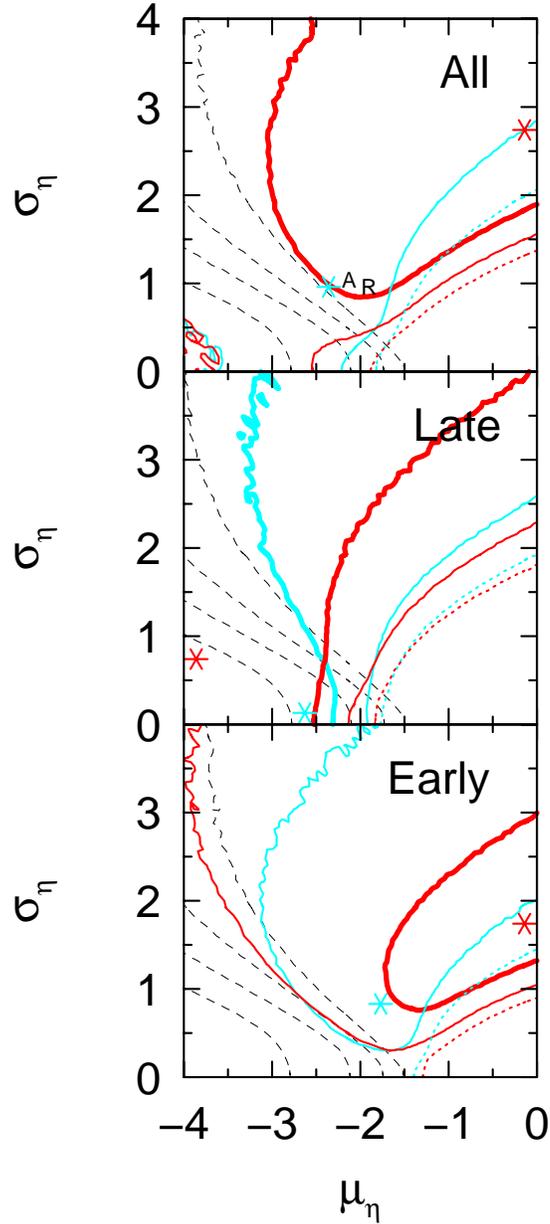}
\caption{\emph{Top}: Asterisks indicate the best fitting values of
$\mu$ and $\sigma$ to the complete sample of LGA spirals, assuming
a lognormal distribution of disk ellipticities. The heavy solid
line, light solid line, and dotted line indicate the $P = 0.5$,
$P = 0.1$, and $P = 0.01$ isoprobability contours.
Red = shapes deduced from $K_s$ data; blue = shapes
from $B$ data. Dashed lines show the scatter in the Tully-Fisher
relation; contours are drawn, starting at the lower left, at 
0.25, 0.5, 0.75, and 1.0 magnitude.
\emph{Middle}: Same, but for spirals of Hubble type Sc or later.
\emph{Bottom}: Same, but for spirals of type Sbc or earlier.
`A' = best fit to data of \citet{an01}; `R' = best fit of \citet{ry04}.
}
\label{fig:lognorm_non}
\end{figure}

In Figures~\ref{fig:gaus_non} and \ref{fig:lognorm_non}, the
dashed lines indicate the expected amount of scatter in the
Tully-Fisher relation if the potential ellipticity is given
by either a Gaussian distribution (Figure~\ref{fig:gaus_non})
or by a lognormal distribution (Figure~\ref{fig:lognorm_non}).
Starting at the lower left corner of each panel, the contours
are drawn at the levels 0.25 mag, 0.5 mag, 0.75 mag, and 1.0 mag.
The inclinations are assumed to be estimated from the apparent
axis ratio. Since the disks are not always infinitesimally thin,
the inclination $\theta_{\rm phot}$ is estimated using the relation
\begin{equation}
\cos^2 \theta_{\rm phot} = {q^2 - \mu_\gamma^2 \over 1 - \mu_\gamma^2} \ .
\end{equation}
Ultrathin galaxies with $q < \mu_\gamma$ are assumed to have
$\cos \theta = 1$. All galaxies with $\theta_{\rm phot} < 45^\circ$
are discarded.

If all spirals are considered together, the best fitting
ellipticity distributions -- either Gaussian or lognormal -- imply
approximately 1 mag of scatter if the potential ellipticity
equals the disk ellipticity: this can be seen in the upper panels 
of Figures~\ref{fig:gaus_non} and \ref{fig:lognorm_non}. The best
fitting ellipticity for the late spirals, as seen in the middle panels
of Figures~\ref{fig:gaus_non} and \ref{fig:lognorm_non}, implies
only 0.3 mag of scatter. Finally, the best fitting ellipticity for
the early spirals, seen in the lower panels of Figures~\ref{fig:gaus_non}
and \ref{fig:lognorm_non}, leads to 1.4 mag of scatter in the
Tully-Fisher relation.
By comparing the predicted scatter
from disk ellipticity to the actual scatter in the Tully-Fisher
relation for these galaxies, I can place constraints on how much
of the scatter can be attributed to the disk ellipticity.

To create a Tully-Fisher plot for the spirals in the 2MASS LGA
sample, I took the relevant astrophysical parameters from the
HyperLeda database.\footnote{
See http://leda.univ-lyon1.fr/}
As a measure of galaxy luminosity, I used the $I$ band absolute
magnitude; multiband studies of the Tully-Fisher relation
indicate that the observed scatter is minimized in or near
the near-infrared.  The $I$ band absolute magnitude is
computed from the apparent magnitude $m_I$ (corrected
for galactic extinction and internal extinction) and the distance
modulus $m-M$. The distance modulus was computed from the radial velocity
$v_{\rm Vir}$ corrected for infall of the Local Group toward Virgo,
assuming a Hubble constant $H_0 = 70 {\rm\,km}{\rm\,s}^{-1}
{\rm\,Mpc}^{-1}$; for galaxies with $v_{\rm Vir} < 500 {\rm\,km}
{\rm\,s}^{-1}$, the distance modulus was taken from the literature,
with preference given to Cepheid distances. As a measure of rotation
speed, I used the maximum rotation velocity $v_m$ determined from
the 21 cm line of neutral hydrogen. In computing $v_m$, the inclination
used is the photometric estimate $\theta_{\rm phot}$, using the
axis ratio of the $25 {\rm\,mag}{\rm\,arcsec}^{-2}$ isophote as the
apparent axis ratio $q$.

\begin{figure}
\plotone{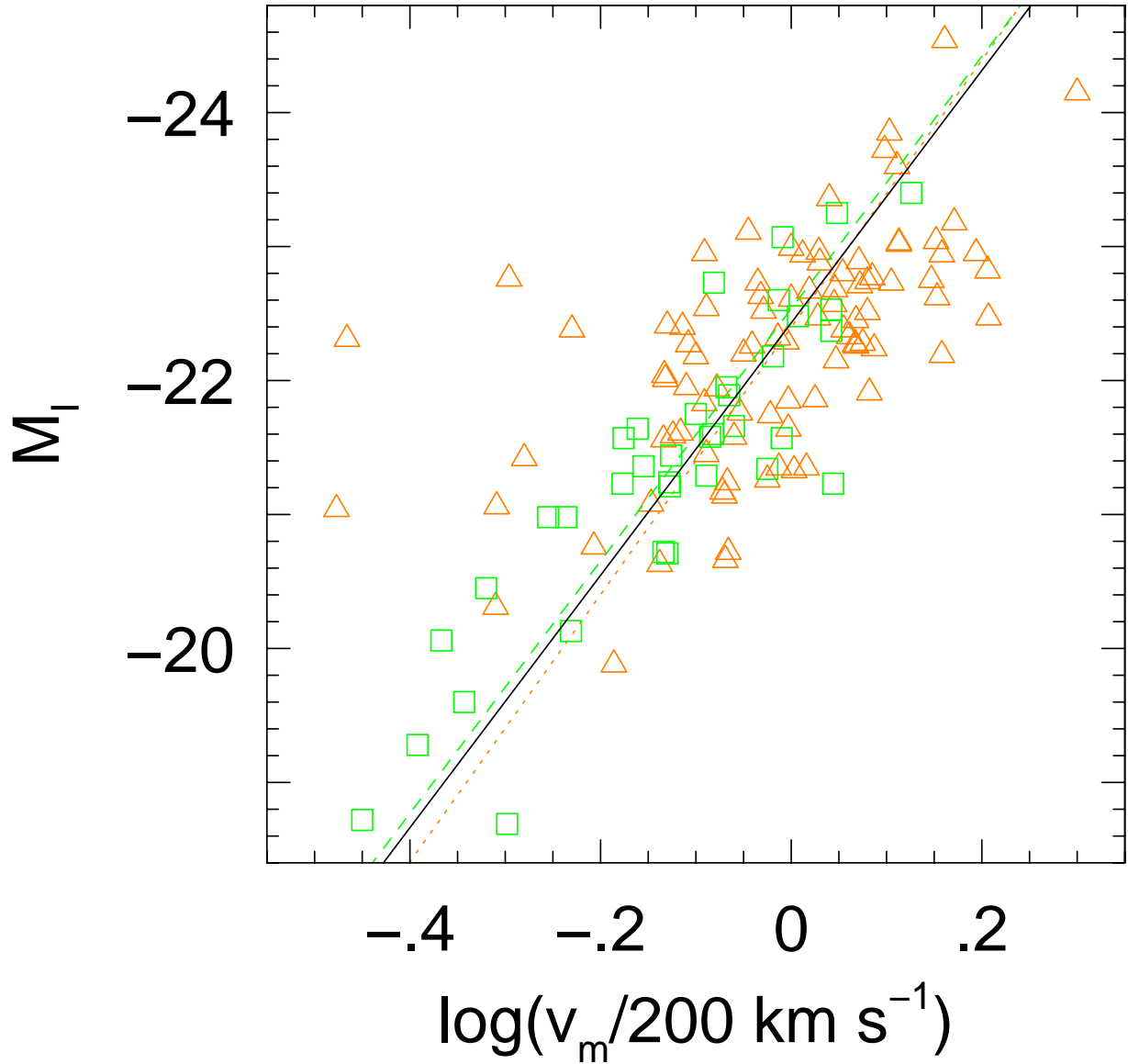}
\caption{The corrected $I$ band absolute magnitude versus the
logarithm of the maximum 21 cm rotation velocity. Squares indicate
late spirals; triangles indicate early spirals. The solid line
is the inverse unweighted fit to all the points ($n = 128$),
the dashed line is the fit to the late spirals only ($n = 37$),
and the dotted line is the fit to the early spirals ($n = 91$).
}
\label{fig:tf}
\end{figure}

Figure~\ref{fig:tf} shows $M_I$ as a function of $\log v_m$ for the 2MASS
LGA spirals with $R_{20}^o > 60 \arcsec$, assuming axisymmetry; galaxies
with $\theta_{\rm phot} < 45^\circ$ are omitted, as are galaxies without
$I$ band photometry in HyperLeda. A total of $n = 128$ galaxies meet
all these criteria. The solid line in Figure~\ref{fig:tf} is the best
fitting Tully-Fisher relation, found using an unweighted inverse
fit \citep{sc80,tu00,ka02}. The best Tully-Fisher relation for the
sample of $n = 128$ galaxies is
\begin{equation}
M_I ( {\rm all} ) = (-22.43 \pm 0.09) - (9.42 \pm 0.77) \log_{10} \left(
{v_m \over 200 {\rm\,km}{\rm\,s}^{-1} } \right) \ ,
\end{equation}
with a scatter in $M_I$ of $\sigma_{\rm rms} = 0.94$ mag.
If only the $n = 37$ late spirals
are considered, an unweighted inverse fit yields
\begin{equation}
M_I ( {\rm late} ) = (-22.54 \pm 0.15 ) - ( 9.42 \pm 0.85 ) \log_{10} \left(
{v_m \over 200 {\rm\,km} {\rm\,s}^{-1} } \right) \ ,
\end{equation}
with a scatter in $M_I$ of $\sigma_{\rm rms} = 0.62$ mag.
Finally, if only the $n = 91$
early spirals are considered, the fit is
\begin{equation}
M_I ( {\rm early} ) = ( -22.40 \pm 0.12 ) - ( 9.97 \pm 1.38 ) \log_{10}
\left( {v_m \over 200 {\rm\,km} {\rm\,s}^{-1} } \right) \ ,
\end{equation}
with a scatter in $M_I$ of $\sigma_{\rm rms} = 1.10$ mag.
The late and early type spirals have
Tully-Fisher relations with statistically indistinguishable slopes;
however, the early spirals have a much greater scatter.

The observed scatter, $\sigma_{\rm rms}$, is not due entirely to
the nonaxisymmetry of disks. In addition to an intrinsic scatter
$\sigma_{\rm ell}$, due to the ellipticity of the disk and the
potential, there is also a contribution $\sigma_{\rm err}$ due
to errors in observation and interpretation. For instance, the
absolute magnitude $M_I$ is subject to errors in flux measurement,
errors in determining distance, and errors in extinction correction.
The rotation speed $v_m$ is subject to errors in measuring 21 cm
line widths and errors in converting line widths to deduced
rotation speeds. Let us suppose that the two sources of scatter
add in quadrature: $\sigma_{\rm rms}^2 = \sigma_{\rm ell}^2 +
\sigma_{\rm err}^2$. If $\sigma_{\rm err}$ is the same for both
early and late spirals, then
\begin{equation}
\sigma_{\rm ell} ({\rm early})^2 - \sigma_{\rm ell} ({\rm late})^2
= \sigma_{\rm rms} ({\rm early})^2 - \sigma_{\rm rms} ({\rm late})^2
= ( 0.91 {\rm\,mag})^2 \ .
\end{equation}
If late spirals are perfectly axisymmetric, which is permitted
by the data, then $\sigma_{\rm ell} ({\rm late}) = 0 {\rm\,mag}$
and $\sigma_{\rm ell} ({\rm early}) = 0.91 {\rm\,mag}$. The
largest possible values of $\sigma_{\rm ell}$ come if I
na{\"\i}vely assume that $\sigma_{\rm err} = 0$; in this
limiting case, $\sigma_{\rm ell} ({\rm late}) = 0.62 {\rm\,mag}$
and $\sigma_{\rm ell} ({\rm early}) = 1.10 {\rm\,mag}$.

For late spirals, the Tully-Fisher relation thus implies that
the intrinsic scatter due to ellipticity lies in the range
$\sigma_{\rm ell} = 0 \to 0.62 {\rm\,mag}$. This is consistent
with the value $\sigma_{\rm ell} = 0.3 {\rm\,mag}$ derived
from the best fitting distributions of disk ellipticity.
For early spirals, the Tully-Fisher relation implies an
intrinsic scatter due to ellipticity in the range
$\sigma_{\rm ell} = 0.91 \to 1.10 {\rm\,mag}$. This is smaller
than the value $\sigma_{\rm ell} = 1.4 {\rm\,mag}$ derived
from the best fitting distributions of disk ellipticity.
However, in the lower panels of Figures~\ref{fig:gaus_non}
and \ref{fig:lognorm_non},
the band where $\sigma_{\rm ell} = 0.91 \to 1.10$ overlaps
the region where $P > 0.1$; thus, the ellipticities derived
from the apparent shapes and from the Tully-Fisher relation
are not discrepant at a high confidence level.

The difference in disk ellipticity between earlier and later
spirals, and the resulting difference in Tully-Fisher scatter,
helps to explain the apparent discrepancy between the relatively
large disk ellipticity ($\epsilon \sim 0.1$) seen in photometric
studies \citep{bi81,gr85,la92,fa93,al02,ry04} and the relatively
small intrinsic scatter ($\sigma \lesssim 0.3 {\rm\,mag}$) seen in
some studies of the Tully-Fisher relation. For instance,
\citet{ve01}, in his study of spirals in the Ursa Major
cluster, found that his DE (distance estimator) subsample
of spirals had a Tully-Fisher relation with 0.26 magnitudes
of scatter in the $I$ band; the intrinsic scatter had a most
likely value of only 0.06 magnitudes. This small scatter
was partly due to the strict selection criteria imposed;
galaxies were excluded if they were obviously interacting,
had prominent bars, or had irregular outer isophotes. Galaxies
of Hubble types earlier than Sb or later than Sd were also
excluded. The scatter was further minimized by the fact that
the resulting subsample of 16 spirals contained 10 galaxies
of type Sc, Scd, and Sd (late spirals, in my classification)
and only 6 galaxies of type Sb and Sbc. If the spirals in the
2MASS LGA are characteristic of all spiral galaxies, then
scatter in the Tully-Fisher relation can be reduced by excluding
spirals of Hubble type earlier than Sc.

In addition to the practical implications for minimizing scatter
in the Tully-Fisher relation, the difference in ellipticity
between early and late spirals provides an intriguing clue
for the formation and evolution of spiral galaxies. Why should
the Sb and Sbc galaxies dominating the early spiral sample be
more elliptical at $R_{25}$ than the Sc galaxies? One possibility
is that the triaxiality of a dark halo's potential affects
the Hubble type of the spiral embedded within it. \citet{ko04}
estimate that the majority of spirals with Hubble type of
Sb and later contain ``pseudobulges'' rather than classical
bulges. Pseudobulges are formed as gas is transported to
small radii by nonaxisymmetric structures such as triaxial
halos, spiral structure, elliptical disks, and bars. Simulations
of disk galaxies in triaxial dark halos \citep{el98,el01}
indicate that pseudobulges grow by secular evolution on a timescale
of a few Gyr. The Hubble type of the resulting galaxy depends
on both the halo core radius and the potential asymmetry, with
greater halo nonaxisymmetry leading to larger bulges.

High resolution n-body simulations of virialized dark halos in
the mass range $10^{12} {\rm\,M}_\odot \lesssim M \lesssim
10^{14} {\rm\,M}_\odot$ indicate that virialized halos are
well described as triaxial shapes \citep{ji02,al05}. More
massive halos are further from spherical, on average. In a
lambda CDM simulation, the mean short-to-long axis ratio
of a halo with mass $M$ at the present day ($z = 0$)
is $\langle \gamma \rangle \approx 0.54 ( M / M_* )^{-0.05}$,
where $M_* = 1.2 \times 10^{13} {\rm\,M}_\odot$ is the characteristic
nonlinear mass scale \citep{al05}. The mean intermediate-to-long
axis ratio also decreases with increasing mass, going from 
$\langle \beta \rangle \approx 0.76$ at $M = 0.1 M_*$ to
$\langle \beta \rangle \approx 0.71$ at $M = M_*$. (Dissipation
by baryons tends to compress the dark halo into a more nearly
spherical shape in its central regions \citep{ka04}, but
will not destroy the trend that lower-mass halos are rounder
on average in the plane of the baryonic disk.) At a given
halo mass, there exists a fairly wide spread in $\beta$ around
the mean value, with standard deviation $\sigma_\beta \sim 0.1$
for $\langle \beta \rangle \sim 0.7$ \citep{al05}. The nearly
axisymmetric halos will give rise to nearly circular disks and
smaller pseudobulges, such as those seen in Sc galaxies. The
more nonaxisymmetric halos will give rise to elliptical disks
and larger pseudobulges, such as those seen in Sb and Sbc galaxies.
The mean mass of Sb spirals is greater than that of Sc spirals;
however, the spread in masses for a given Hubble type is greater
than the difference in the mean \citep{ro94}. This would be
expected if the difference between early (Sb and Sbc) spirals
and late (Sc) spirals is not due primarily to differences in
halo mass, but rather is due to differences in halo nonaxisymmetry.

\acknowledgments

This publication makes use of data products from the Two Micron
All Sky Survey, which is a joint project of the University of
Massachusetts and the Infrared Processing and Analysis Center /
California Institute of Technology, funded by the National
Aeronautics and Space Administration and the National Science
Foundation. It also makes use of the HyperLeda galaxy database.
James Pizagno provided useful Tully-Fisher insights.


\begin{deluxetable}{llcccccccc}
\tablewidth{0pt}
\tablecaption{Best Parametric Fits\label{tab:param}}
\tablehead{
\colhead{Sample} & \colhead{Band} & \colhead{$\mu_\gamma$} & 
\colhead{$\sigma_\gamma$} & \colhead{ } & \colhead{$\mu_\epsilon$} & 
\colhead{$\sigma_\epsilon$} &  \colhead{ } & \colhead{$\mu_\eta$} & \colhead{$\sigma_\eta$} }
\startdata
All   & $K_s$ & 0.26 & 0.09 & & 0.01 & 0.26 & & -0.14 & 2.74 \\
      &  $B$  & 0.17 & 0.03 & & 0.01 & 0.16 & & -2.36 & 0.96 \\
Late  & $K_s$ & 0.18 & 0.06 & & 0.02 & 0.02 & & -3.86 & 0.74 \\
      &  $B$  & 0.14 & 0.02 & & 0.07 & 0.02 & & -2.63 & 0.13 \\
Early & $K_s$ & 0.29 & 0.08 & & 0.02 & 0.47 & & -0.14 & 1.74 \\
      &  $B$  & 0.20 & 0.04 & & 0.09 & 0.22 & & -1.77 & 0.83 \\
\enddata
\end{deluxetable}

\end{document}